\documentclass[fleqn,usenatbib]{mnras}
\pdfoutput=1

\usepackage[T1]{fontenc}
\usepackage{ae,aecompl}

\usepackage{graphicx}	% Including figure files
\usepackage{amsmath}	% Advanced maths commands
\usepackage{amssymb}	% Extra maths symbols
\usepackage{longtable}
\usepackage{xspace}
\usepackage{bm}

\newcommand{\Msun}{M$_{\odot}$\xspace}
\newcommand{\Mvir}{M$_{\rm vir}$\xspace}

\newcommand{\Rvir}{R$_{\rm vir}$\xspace}

\newcommand{\kms}{km s$^{-1}$\xspace}
\newcommand{\tcross}{t$\rm_{cross}$\xspace}

\title[Masses of the MW and M31]{Orbits of massive satellite galaxies -- II. Bayesian Estimates of the Milky Way and Andromeda masses using high
precision astrometry and cosmological simulations}

\author[E. Patel et al.]{
Ekta Patel,$^{1}$\thanks{E-mail: ektapatel@email.arizona.edu}
Gurtina Besla,$^{1}$
and Kaisey Mandel$^{2}$
\\ 
$^{1}$Department of Astronomy, University of Arizona, 933 North Cherry Avenue, Tucson, AZ 85721, USA\\
$^{2}$Harvard-Smithsonian Center for Astrophysics, 60 Garden Street, Cambridge, MA 02138, USA\\}

\date{Accepted XXX. Received YYY; in original form ZZZ}

\pubyear{2016}

\begin{document}
\label{firstpage}
\pagerange{\pageref{firstpage}--\pageref{lastpage}}
\maketitle

% Abstract of the paper (250 words max)
\begin{abstract}

In the era of high precision astrometry, space observatories like the {\em Hubble Space Telescope} (HST) and {\em Gaia} are providing unprecedented 6D phase space information of satellite galaxies.  Such measurements can shed light on the structure and assembly history of the Local Group, but improved statistical methods are needed to use them efficiently. Here we illustrate such a method using analogs of the Local Group's two most massive satellite galaxies, the Large Magellanic Cloud (LMC) and Triangulum (M33), from the Illustris dark-matter-only cosmological simulation. We use a Bayesian inference scheme combining measurements of positions, velocities, and specific orbital angular momenta (\;{\em j}\;) of the LMC/M33 with importance sampling of their simulated analogs to compute posterior estimates of the Milky Way (MW) and Andromeda's (M31) halo masses. We conclude the resulting host halo mass is more susceptible to bias when using measurements of the current position and velocity of satellites, especially when satellites are at short-lived phases of their orbits (i.e. at pericentre). Instead, the {\em j} value of a satellite is well-conserved over time and provides a more reliable constraint on host mass. The inferred virial mass of the MW (M31) using {\em j} of the LMC (M33) is $\rm M_{vir, MW}=1.02^{+0.77}_{-0.55}\times10^{12}$ \Msun ($\rm M_{vir, M31}=1.37^{+1.39}_{-0.75}\times10^{12}$ \Msun). Choosing simulated analogs whose {\em j} values are consistent with the conventional picture of a previous (< 3 Gyr ago), close encounter (< 100 kpc) of M33 about M31 results in a very low virial mass for M31 ($\sim\!10^{12}$ \Msun). This supports the new scenario put forth in \citet{patel16}, wherein M33 is on its first passage about M31 or on a long period orbit. We conclude that this Bayesian inference scheme, utilising satellite {\em j}, is a promising method to reduce the current factor of two spread in the mass range of the MW and M31. This method is easily adaptable to include additional satellites as new 6D phase space information becomes available from HST, {\em Gaia} and JWST.
\end{abstract}

\begin{keywords}
Galaxy: fundamental parameters -- galaxies: evolution -- galaxies: kinematics and dynamics -- Local Group
\end{keywords}

%%%%%%%%%%%%%%%%%%%%%%%%%%%%%%%%%%%%%%%%%%%%%%%%%%

%%%%%%%%%%%%%%%%% BODY OF PAPER %%%%%%%%%%%%%%%%%%

\section{Introduction}
\label{sec:intro}
The Milky Way (MW) and Andromeda (M31) both host a plethora of known substructures within their respective dark matter haloes. These substructures include dwarf satellite galaxies, globular clusters, and also several stellar streams. Most of these systems are dynamically bound to their given host halo at present, making them unique tracers of their host's gravitational potential. 

With instruments like the {\em Hubble Space Telescope} (HST), proper motions of dwarf satellite galaxies, globular clusters, and stellar streams have been measured with microarcsecond per year precision. High precision astrometry promises to be an especially fruitful field with the recent launch of the {\em Gaia} satellite. Results from {\em Gaia} data release 1 \citep[DR1,][]{gaia16b} have already confirmed the proper motions of the Magellanic Clouds \citep{vdm16} previously measured by \citet[][hereafter K13]{k13} and others \citep[e.g.,][]{k06a, k06b, piatek08, vieira10}. We aim to leverage the full 6D phase space information for nearby, massive satellite galaxies ($\sim$$10^{11}$ \Msun) to inform us about the assembly and structure of our Local Group of galaxies using high resolution dark matter simulations of large cosmological volumes.  

The precise motion of satellite galaxies and remnant streams have already acted as a stepping stone for dynamical mass estimates of the MW. Their 3D positions and velocities derived from the proper motions are used as instantaneous tracers of the halo potential and can therefore estimate the total mass enclosed within a given radius. For example, numerical models designed to reproduce properties of the stellar debris in the Sagittarius stellar stream have yielded conflicting estimates on the mass of the MW. Estimates derived from the tidal disruption of the Sagittarius dSph imply a rather light MW mass of M(200 kpc)=$5.6\pm1.2\times10^{11}$ \Msun \citep[68 per cent credible interval;][]{gibbons14}, but more recent simulations of the Sagittarius stellar stream are able to reproduce its kinematics in a MW mass of order $10^{12}$ \Msun within 206 kpc \citep{dierickx16}. Such orbital models have not yet converged to a consistent result for the mass of the MW and demonstrate just one instance of ambiguity in its measurement. 

Many other independent methods have also been used to measure the mass of the MW. \citet{zaritsky89, kochanek96, wilkinson99, sakamoto03, eadie17} have all considered the motion of multiple satellite galaxies, globular clusters, or both to determine the MW's mass. Other methods include abundance matching between cosmological simulations and observational surveys \citep[e.g.][]{moster13}, applying the cosmological baryonic fraction of the MW to estimate the lower bound on its mass without invoking dynamics \citep{zaritsky17}, computing mass via the MW-M31 timing argument \citep[e.g.][]{vdm12ii, vdm12iii}, and more \citep[see][]{dehnen98, dehnen06, moore94, murali00, binney01, rasmussen01, klypin02, smith06, brown06}. 

Cosmological simulations provide an independent statistical method for constraining the MW's halo mass under the assumption that they accurately capture the physics and underlying cosmology of our Universe. Together, with high precision astrometry, these simulations have opened a new door for near--field cosmology. Dynamical properties, such as orbital energy and angular momentum, computed from 6D phase space measurements can be used to statistically infer the total mass of a host galaxy's halo. \citet[][hereafter BK11]{bk11} used the dynamics of the Magellanic Clouds (MCs) from the \citet{k06a,k06b} proper motions and the frequency of their analogs in the Millennium-II Simulation \citep{bk09} to conclude that the MW's virial mass is $\geq 2\times10^{12}$ \Msun. In \citet[][hereafter Paper I]{patel16}, we followed a similar methodology using revised proper motions of the LMC from K13 and the Illustris dark matter-only cosmological simulation \citep{nelson15, vogelsberger14B, vogelsberger14A} to illustrate that the hosts of LMC-like systems (of similar mass and orbital energy) have typical halo masses of order $1.5\times10^{12}$ \Msun. 

A similar analysis can now be applied to M31 for the first time, as its proper motion was only recently measured with HST \citep[][hereafter S12 and vdM12]{sohn12, vdm12ii}. The proper motion for M33, the most massive satellite galaxy of M31, was measured by observing water masers with the {\em Very Long Baseline Array} \citep[][hereafter B05]{brunthaler05}. These combined measurements enable us to study both the MW-LMC and M31-M33 systems as isolated host-satellite systems in tandem. In Paper I, we demonstrated that satellites identified in Illustris with masses and specific orbital energy comparable to that of M33 are most likely on their first approach about their hosts. The M31 analogs that host such satellites have typical halo masses $\geq1.5\times10^{12}$ \Msun. Many independent efforts have also been made to estimate the mass of M31 \citep[e.g.][]{klypin02, watkins10, tollerud12}. 

While the above numerical and cosmological methods are promising, the MW's plausible mass range is $\approx 0.7-1.5 \times 10^{12}$ \Msun and that of M31's is $\approx 1.5-2.5 \times 10^{12}$ \Msun. Observational evidence shows that the total mass of M31 should be higher than that of the MW's as M31's stellar disk is more massive and it hosts dwarf elliptical galaxies. We will demonstrate that inferred masses of the MW and M31 using only positions and velocities of satellites contradict this general belief. 

The advent of high mass resolution cosmological simulations with large volumes ($\gtrsim$100 Mpc per side; e.g. Illustris, EAGLE, Millennium-II, Bolshoi) has provided a statistically significant data set to explore a novel inference method that may help us to further constrain this mass range range for the MW \citep[][hereafter B11]{busha11} and for the Local Group \citep[][hereafter G13]{gonzalez13}. B11 developed and applied a Bayesian inference scheme to a set of Magellanic Cloud (MC) analogs in the Bolshoi \citep{klypin11} cosmological simulation using the observed positions, velocities, and circular velocities derived from their proper motions \citep{k06a, k06b, k13}. Assuming that the errors on these measured properties are Gaussian, they invoked an importance sampling technique to infer the posterior distribution of the MW's halo mass. 

One major assumption in all of these studies that utilised inference techniques is that the position and velocity of the LMC today are typical, however, it is well known that the LMC is likely just past pericentre, and such orbital configurations are short-lived \citep{b07,besla12, k13, gomez15}. Furthermore, the position and velocity of the LMC today are rare amongst the phase space of known Local Group dwarf satellites. In contrast, M33 is between apo- and pericentre, and therefore exhibits a less transient configuration \citep{patel16, putman09, mcconnachie09}. 

G13 examined the effects of the larger environment of the MW (i.e. including an M31 companion galaxy) in determining its mass and found MW mass estimates in agreement with B11,  concluding that the requirement of a Local Group environment does not affect the inferred mass of the MW. More recent work (Williamson et al., in prep.) uses the combined constraints from the MW--LMC--M31--M33 to identify analogs of the Local Group and place further constraints on these mass estimates with a Bayesian approach. \citet{carlesi16} have also obtained the mass of the Local Group in a statistical fashion, finding a MW mass estimate of 0.6-0.8$\times10^{12}$ \Msun, somewhat lower than that typically determined with the timing argument. 

In this work, we will focus on the specific orbital angular momentum of the LMC (and M33), as it is generally well conserved with time, and use it to infer the most typical MW (M31) mass. By doing so, we aim to avoid any bias that may be introduced due to the transient nature of the LMC's current orbital configuration. Using these two massive satellites galaxies in tandem to constrain their respective host halo masses will test the robustness of the adopted Bayesian inference technique while also providing insight on how the orbital histories of massive satellites can uncover important properties of their host environment.

For this paper, we allow the halo mass of the host galaxy (the MW or M31) to be a free parameter and estimate its most probable value using the present-day dynamics of the LMC or M33 in combination with the Illustris cosmological simulation via Bayesian methods adopted from B11. Recent proper motion measurements and the higher mass resolution of Illustris motivate us to re-examine the MW-LMC system. While \citet{fardal13} have inferred the mass of M31 in a Bayesian fashion using constraints from the Giant Southern Stream, we extend B11's Bayesian method to compute the mass of M31 using observed properties of M33 for the first time. In the era of high astrometric precision, these types of statistical analyses will be key to refining our understanding of the Local Group. In future work, we will further explore how this technique may be expanded to include more (less massive) satellite galaxies as their proper motions are obtained with HST and {\em Gaia} in the upcoming years.

This paper is organized as follows. In Section~\ref{sec:illustris}, we describe the dark matter-only Illustris cosmological simulation and the sample criteria for identifying a control set of host-satellite pairs analogous to the MW-LMC or M31-M33. Section~\ref{sec:bayesian} details the Bayesian inference method implemented to determine the posterior mass distributions for the MW and M31. In Section~\ref{sec:results}, we present results for the masses of the MW and M31 using two different likelihoods in combination with the properties of the LMC and M33, respectively. Section~\ref{sec:discussion} further discusses the implications of different satellite orbital histories on the mass estimates of their host galaxies, the impact of measurement and cosmic variance errors on this analysis, and the results of this method as compared to previous work. Finally, Section~\ref{sec:conclusions} contains a summary of our findings and addresses future prospects.
 
\section{The Illustris Simulation and Sample Selection} 
\label{sec:illustris}

In Paper I of this series, we identified several hundred massive satellite analogs of the LMC and M33 in the {\em Illustris-1-Dark} (hereafter Illustris-Dark) cosmological simulation \citep{nelson15, vogelsberger14B, vogelsberger14A}. We found that orbital energy shows a tight correlation with host halo mass \citep[see also][]{bk11}. As the absolute value of the specific orbital energy increases, the host mass also increases. However, satellites may spend a significant amount of time in orbit about their hosts and suffer orbital decay owing to dynamical friction. As such, here we utilise this control sample of massive satellite analogs to gauge the stability of dynamical properties of satellite orbits, such as orbital angular momentum and orbital energy over time. 
 
Since we use analogs of the LMC and M33, which are currently at different positions within their orbits, we must first identify properties of these satellites that remain stable with time so that our analysis is consistent for both host-satellite systems. 

In the following, we describe the specifications of the Illustris-Dark dark matter-only cosmological simulation and the criteria for selecting a control sample of host-satellite pairs that mimic the mass ratio of the MW-LMC and M31-M33 systems. The host-satellite control sample is used to determine which satellite dynamical properties are most suitable for the statistical analysis described in Section~\ref{sec:bayesian}. 

\subsection{Simulation}
The {\em Illustris Project}\footnote{The Illustris catalogs are all publicly available at www.illustris-project.org} is a suite of N-body+hydrodynamic simulations run with the \texttt{AREPO} code, spanning a cosmological volume of (106.5 Mpc)$^3$ \citep{nelson15, vogelsberger14B, vogelsberger14A, genel14}. As in Paper I, we use only the Illustris-Dark simulation,  which follows the evolution of 1820$^3$ dark matter particles from $z=127$ to $z=0$. Illustris-Dark uses the {\em WMAP-9} cosmological parameters \citep{hinshaw13}:

\begin{multline} \Omega_m =0.2726,~~~~\Omega_{\Lambda} = 0.7274,~~~~\Omega_b =0.0456, \\*
\!\!\!\! \sigma_8 = 0.809,~~~~n_s = 0.963,~~~~h = 0.704. \end{multline} 
These cosmological parameters differ slightly from the parameters used in the Bolshoi \citep{klypin11} and Millennium-II \citep{bk09} cosmological simulations. However, we have reproduced the methodology of previous studies, as described later, and recover consistent results. We will make further comparisons to previous work in Section~\ref{subsec:previouswork}.

Haloes and halo substructure in Illustris-Dark are identified with the \texttt{SUBFIND} routine \citep{springel01, dolag09}. We use the Illustris-Dark merger trees created with the \texttt{SUBLINK} code \citep{rg15} to trace the orbital histories of massive satellite analogs in this analysis. In addition to the full orbital histories, merger trees also provide information about the mass and size evolution of both hosts and satellites throughout cosmic time.

The Bolshoi simulation has a much larger simulation volume compared to Illustris-Dark (250 $h^{-1}$ Mpc per side vs. 75 $h^{-1}$ Mpc per side), however the dark matter particle mass is only of order $10^{8}$ \Msun. Upon identifying LMC/M33 mass analogs by our definition (see Paper I) in the Bolshoi simulation, each analog would only consist of $10^{2}-10^{3}$ dark matter particles, whereas Illustris provides at least $10^{3}-10^{4}$ dark matter particles per massive satellite analog with a dark matter mass resolution of $\rm m_{DM}=7.5 \times 10^6$ \Msun. Thus, while Bolshoi will provide a larger statistical sample of massive satellite analogs, the Illustris-Dark analogs are individually better resolved. 
\subsection{Control Sample Selection}
\label{subsec:control}
MW/M31 analogs are all central subhaloes (i.e. the primary subhalo containing the majority of the bound material in a given halo as determined by \texttt{SUBFIND}) whose halo virial mass at $z=0$ is \Mvir$=0.7-3\times 10^{12}$ \Msun. We use this generous mass range to reflect all reported values for masses of the MW and M31 in recent literature. In total, 1933 haloes satisfy these criteria. Therefore, MW/M31 mass analogs are composed of order $10^5$ dark matter particles each. Virial mass and virial radius for all MW/M31 analogs are taken directly from the Illustris-Dark halo catalogs and are based on the spherical tophat approximation. 

Host haloes are then defined as the subset of MW/M31 analogs which also host a massive subhalo like the LMC or M33. Section 5 of Paper I outlines more details regarding the sample selection criteria of host haloes and their subsequent massive satellite analogs (also see BK11). In Illustris-Dark, we find about 24.4 per cent of MW/M31 mass halo analogs host a massive satellite analog like the LMC or M33 within their virial radius. This frequency is consistent with observational surveys of L$_*$ galaxies and previous theoretical studies of the MCs using cosmological simulations \citep[e.g.][]{tollerud11,bk11, liu11}. The full control sample of host-satellite analogs consists of 472 systems. See Figs. 3 and 4 in Paper I for the distribution of host halo virial mass and the host to satellite mass ratios. This sample of host-satellite pairs will only be used as a control sample in this paper, specifically to test the stability of satellite orbital dynamics over time in the following section.

\subsubsection{The Evolution of Satellite Orbital Dynamics in Illustris}
\label{subsubsec:dynamics}

\begin{figure*}
\begin{center}
\includegraphics[scale=0.8]{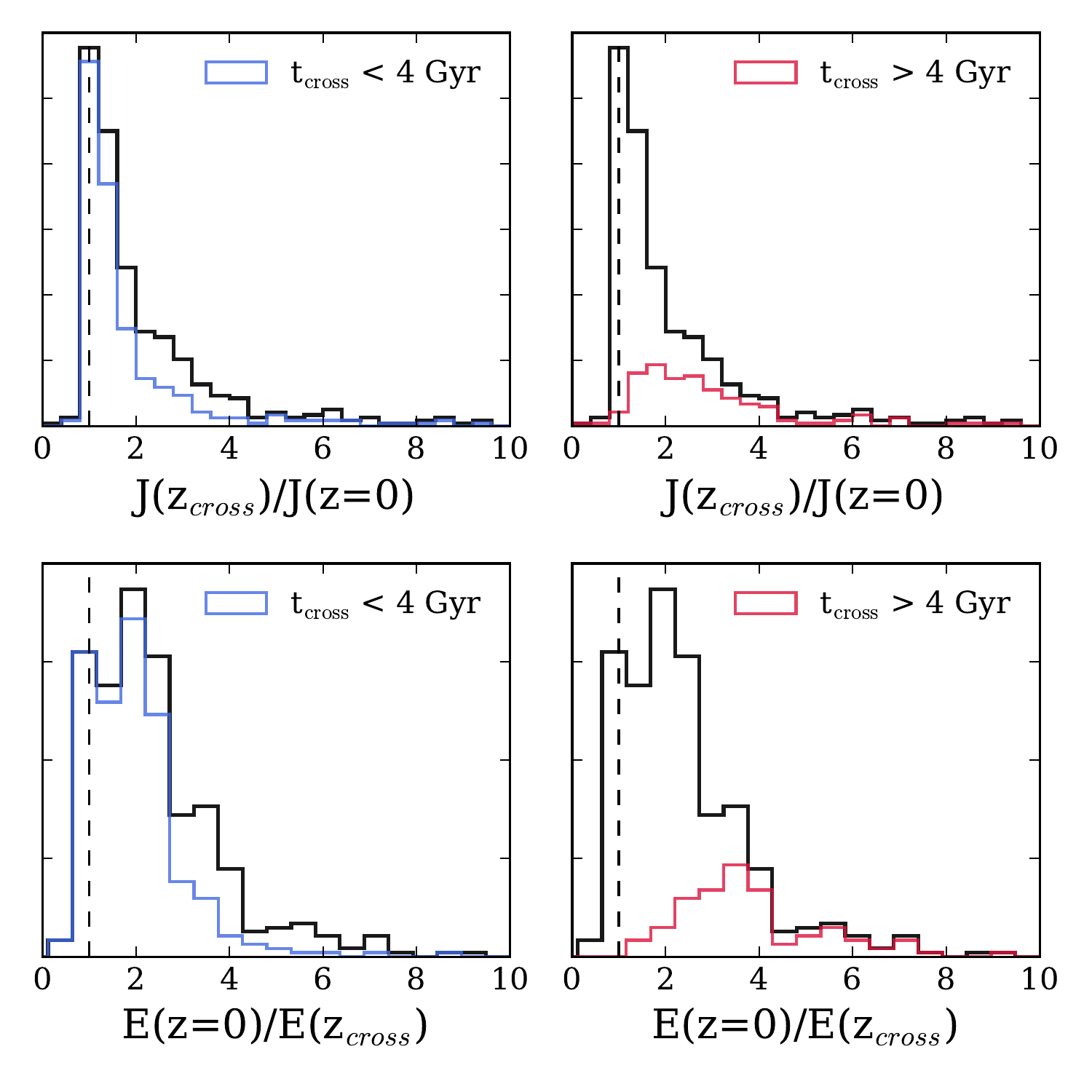}
\caption{{\em Top:} The ratio of specific orbital angular momentum for the Illustris-Dark massive satellite control sample with \tcross $<$ 4 Gyr (left panel, blue) and the \tcross $>$ 4 Gyr satellite sample (right panel, red). The orbital angular momentum ratios peak around one (vertical dashed line) for both samples indicating very little angular momentum loss between infall and $z=0$. {\em Bottom:} The ratio of specific orbital energy for the the \tcross $<$ 4 Gyr sample (left panel, blue) and the \tcross $>$ 4 Gyr sample (right panel, red). Orbital energy varies more significantly and changes by up to a factor of four for most systems. For reference, the distribution of angular momentum ratios and energy ratios for the entire sample are given in black solid histograms. All histograms are normalised to the size of the full sample (black) such that each pair of blue and red histograms sums to the black histogram when stacked.}
\label{fig:dynamicratios}
\end{center}
\end{figure*}

Previous studies have estimated the dark matter halo mass of the MW based on the observed properties of the MCs, such as relative position, velocity, and maximum circular velocity (BK11, B11, G13). For example, the properties of the LMC used in B11 are: $\rm r^{obs}=50\pm2$ kpc, v$\rm_{tot}^{obs}=378\pm36$ \kms, and v$\rm_{max}^{obs}=65\pm15$ \kms based on the LMC proper motions given by \citet{k06a, k06b}.

Folding these properties into a Bayesian scheme automatically assumes that these observed properties are typical amongst the population of massive satellite galaxies in a given redshift range. It is well known that while the orbits of satellite galaxies can be fairly eccentric upon infall into their host's halo \citep[e.g.][]{wetzel11, benson05}, these orbits decay significantly as they experience dynamical friction and mass loss. Consequently, the positions and velocities of satellites relative to their hosts' motion will evolve accordingly with time. Choosing a satellite based on its instantaneous position and velocity therefore implies a unique location within the orbit, rather than the most {\em typical} location. This can dramatically limit the number of plausible analogs, particularly if the satellite is in an unusual location in its orbit. 

Instead, we examine the total specific orbital angular momentum and the specific orbital energy of massive satellite analogs. By quantifying the time evolution of these orbital `constants', we can assess whether they are accurate tracers of the satellite orbital properties since their time of infall. The infall time is defined as the point in time at which the satellite first crossed their host's time-evolving virial radius (see Paper I, Section 6.1). 

This technique ensures that the massive satellite analogs with orbits most similar to those of the LMC or M33, respectively, are chosen from the Illustris-Dark simulation to estimate the halo mass of the MW or M31. It also eliminates any contamination from satellite analogs that may only instantaneously satisfy a specific position and velocity criteria at $z\approx0$, but which ultimately fail to identify in the same family of orbits as those of interest in this work (i.e. the LMC and M33).

In the following, we compare the stability of specific orbital energy and specific angular momentum for the control sample of host-satellite analogs described above. By doing so, we decipher which quantity is more stable over time, justifying its usage in a Bayesian inference scheme. Specific orbital energy and specific orbital angular momentum for the control sample of massive satellite analogs are calculated using Eqs.~\ref{eq:energy}-\ref{eq:angmom}. 

\begin{equation} \label{eq:energy}  \rm E_{sat} = \frac{1}{2}v^2 + \Phi_{NFW}(M_{vir}, c_{vir}, r) \end{equation} 

In Equation~\ref{eq:energy}, the gravitational potential of each host halo is approximated by a Navarro-Frenk-White \citep[][NFW]{nfw96} profile. The virial concentration, $\rm c_{vir}$, is calculated with the fitting formula of the Bolshoi simulation at $z=0$ \citep{klypin11}:
\begin{equation}  c_{\rm vir}(M_{\rm vir})=9.60\left(\frac{M_{\rm vir}}{10^{12} h^{-1}M_{\odot}}\right). \label{eq:cvir} \end{equation}
The total specific orbital angular momentum is computed by:
\begin{equation} \rm j = | \bf{r} \times \bf{v} |. \label{eq:angmom} \end{equation}
Here {\bf r} is the relative position vector connecting the host and satellite, whereas {\bf v} is the velocity vector of the satellite relative to its host. The total specific orbital angular momentum (j) is therefore the magnitude of their cross product.

For each host-satellite member of our control sample we compute the satellite's specific orbital angular momentum and the specific orbital energy at the redshift of satellite crossing time ($\rm z=z_{cross}$) and at $z=0$. The distribution of the ratio of these values (E($z=0$)/E(z$\rm_{cross}$) and J(z$\rm_{cross}$)/J($z=0$)) is plotted in Fig.\ref{fig:dynamicratios} for all control satellites (black), separated by those accreted at early (t$\rm_{cross} <$ 4 Gyr ago; left) and late (t$\rm_{cross}>$ 4 Gyr ago; right) crossing times\footnote{Infall time is used interchangeably with crossing time--the first time a satellite crosses into the time-evolving virial radius of its host halo.}. The ratios are computed in this order so that the distributions share the same horizontal axis. 

The top panels in Fig.~\ref{fig:dynamicratios} show the distribution of the ratio of specific orbital angular momentum at $z=z_{\rm cross}$ to $z=0$. Recently accreted satellites in the control sample (top left panel) experience an angular momentum change less than a factor of two on average. For the early accreted satellites (top right panel), the angular momentum loss is only slightly more significant, reaching factors of four or six for small fractions of the sample. The latter results are naturally expected because these orbits have decayed more substantially since their time of infall. 

The bottom panels of Fig.~\ref{fig:dynamicratios} highlight that changes in position, velocity, and host halo virial mass result in a loss of orbital energy since the time of infall because satellite orbits decay via dynamical friction (see Paper I, Section 6.3). Dynamical friction is proportional to the satellite mass squared, therefore the more massive the satellite, the faster its orbit decays. Fig.~\ref{fig:dynamicratios} shows that recently accreted satellites can lose up to four times the orbital energy exhibited at infall while early accreted satellites can lose up to eight times their original orbital energy. Generally, the distribution of orbital energy evolution is broader than that of orbital angular momentum, independent of satellite crossing time. This is especially crucial for the most recently accreted satellites since about 70 per cent of the massive satellite analogs in Illustris-Dark were accreted in the last 4 Gyr (see Paper I, Section 6.1).

We conclude that orbital angular momentum is more stable than orbital energy for the population of massive satellite analogs over time. By examining Eqs.~\ref{eq:energy}-\ref{eq:angmom}, it is also clear that orbital angular momentum is not directly correlated with host halo mass as \Mvir does not appear in Eq.~\ref{eq:angmom}, unlike in Eqs.~\ref{eq:energy} and \ref{eq:cvir}. Thus, orbital angular momentum introduces less intrinsic host mass bias. In the following section, total specific orbital angular momentum is treated like an observable to determine the most likely host halo mass for the LMC and M33. This methodology falls in line with action--angle dynamics where actions (angular momentum coordinates) and angles replace position--velocity coordinates in numerically integrated (periodic) orbital models to simplify orbit calculations. For example, \citet{bovy14, sanders14, helmi16, helmi99} track the orbits of various MW substructures using this technique.

This method is of particular interest with regards to the LMC because many previous studies \citep[][B11, K13, and references therein]{patel16,b07} have concluded it is just past pericentre -- a unique epoch in a satellite galaxy's lifetime as it is a short-lived configuration. To constrain the host halo mass with the most physically motivated and informative sample of massive satellites, we must consider the family of orbits to which the LMC belongs rather than just its position and velocity today. 

Note that, in Paper I, we found that M33 could be just past apocentre and therefore more common amongst the phase space of massive satellite analogs. Given this orbital history, it may still be reasonable to consider M33's position and velocity today as an indicator of host halo mass. We will explore both the B11 (position and velocity) likelihood function and a newly developed angular momentum likelihood function in our importance sampling technique moving forwards. 

\section{Bayesian Inference Method}
\label{sec:bayesian}
Now we reverse our analysis from Paper I and constrain host halo mass by using satellite dynamics in a Bayesian inference scheme. The host halo mass is left as a free parameter and is informed only by the observed properties of the LMC/M33 in combination with host-satellite analogs in the Illustris-Dark simulation. The recent HST proper motion analysis of M31 \citep{sohn12, vdm12ii} allows us to apply this method to the M31-M33 system for the first time.

We follow the Bayesian inference method described by B11, who used the halo catalogs from the Bolshoi simulation to estimate the mass of the MW. Note that we focus on the presence of just one massive subhalo analogous to the LMC or M33, while B11 requires each halo to host an analog of both the LMC and SMC. 

The statistical method relies on applying a set of observables as priors to the full Illustris-Dark halo catalog. In the first case, we will examine the resulting masses of the MW and M31 upon considering the position and velocity of the LMC and M33 as independent observables. The second case treats the angular momentum of the satellites as an observable, thereby considering a larger fraction of satellite phase space. The code developed for this work is publicly available on GitHub.

\subsection{Observed Properties}
\label{subsec:observed}

As discussed in Paper I, the proper motions of the LMC, M33 and M31 make it possible to study the orbital histories of the MW-LMC and M31-M33 systems in detail. K13 measured the LMC's proper motions directly using HST, updating previous results from \citet{k06a}. The LMC's proper motions are transformed to Galactocentric positions and velocities using the methods of \citet{vdm02}. Uncertainties on these values are determined by a Monte Carlo scheme that samples the 4$\sigma$ error space of proper motions, radial velocity, position, and the solar motion quantities. This scheme yields 10,000 unique position and velocity vectors from which their standard errors are calculated. These vectors can also be used to compute average dynamical quantities such as orbital angular momentum and its standard error. 

The proper motion of M33 was measured using the {\em Very Long Baseline Array} by B05. M31's proper motion was measured directly, also using HST, by S12. These measurements were corrected for viewing perspective and internal motions by vdM12. Both sets of measurements are transformed to Galactocentric quantities in the same fashion as the LMC. They are combined to yield 10,000 unique position and velocity vectors in the combined error space of the M31-M33 system. Again, these vectors can be used to compute mean position, velocity, and angular momentum of M33 with respect to M31 (see Table 1 of Paper I).

The final observable required for this statistical analysis is the maximum circular velocity of the LMC and M33. Maximum circular velocity is used as a proxy for satellite mass enclosed at a given radius since $\rm v_c^2 = GM_{sat}(r)/r$. See Appendix~\ref{sec:appendixA} for a short discussion on the stability of circular velocity compared to subhalo mass in simulations. 

The LMC's rotation curve was most recently measured by \cite{vdmnk14}, who conclude that its circular velocity peaks at $\rm v_{circ,max} = 91.7\pm18.8$ \kms. M33's rotation curve was measured by \citet{corbellisalucci}; its circular velocity at 15 kpc from its centre (the farthest radial data point measured) is $\rm v_{circ} \sim 130$ \kms, thus we adopt this value for M33's $\rm v_{circ,max}$, although it is expected that the rotation curve continues to rise at larger radial distances. Since we use the dark matter-only version of Illustris throughout this study, we need only to account for the peak circular velocity ($\rm v_{max}^{obs}$) of the dark matter halo. The halo's circular velocity typically peaks in the outer halo where there is minimal contribution from the baryonic disk, which instead peaks within the innermost few kpc of a galaxy.

We adopt the LMC's peak halo velocity modeled by \citet{besla12} and the peak halo velocity of M33 modeled by vdM12. The models estimate the individual contributions of the halo, disc, and bulge for the LMC and M33 such that the total rotation curve reproduces the observed data. We use the peak values of the halo rotation curves in these models and assign a halo peak circular velocity error\footnote{We have also tested an error of 15 \kms (as in B11) and find no significant differences in the results presented in Section~\ref{sec:results}.} of 10 \kms to both satellite velocities \citep[see][]{vdmnk14, corbellisalucci}. The observed properties of the LMC and M33 used in this analysis are summarized in Table~\ref{table:bayesparams}. Note that we have adopted these satellite properties to remain consistent with those used in Paper I. We stress that new measurements of any of these properties can be easily implemented using this methodology (e.g. a more refined measurement of M33's distance\footnote{The 10,000 Monte Carlo samples drawn from the M31-M33 proper motion error space do contain position vectors which reflect the suggested high distance measurement to M33 of $\sim$968 kpc \citep{u09}.}

\begin{table}
\centering
\caption{Observational data ({\bf d}) for the LMC and M33 used to build likelihoods in the Bayesian inference scheme includes the maximum circular velocity, current separation from the host galaxy, and total velocity relative to the host galaxy. a: The maximal circular velocity of the LMC's halo rotation curve is adopted from \protect\citet{besla12}.; b: M33's halo rotation curve maximum is duplicated from \protect\citet{vdm12iii}. M33's position, velocity, and their errors are adopted from Paper I (Table 1) and references within.}
\label{table:bayesparams}
\begin{tabular}{l|c|c||c|c}
& LMC $\mu$ & LMC $\sigma$ & M33 $\mu$ & M33 $\sigma$ \\ \hline
$\rm v_{max}^{obs}$ [\kms]& 85$^{a}$ & 10 & 90$^{b}$ & 10 \\ \hline
$\rm r^{obs}$ [kpc] & 50 & 5 & 203 & 47 \\ \hline 
$\rm v_{tot}^{obs}$ [\kms] & 321 & 24 & 202 & 38 \\ \hline
$\rm j^{obs}$ [kpc \kms] & 15,688 & 1,788 & 27,656 & 8,219 \\ \hline 
\end{tabular}
\end{table}

\subsection{Statistical Methods}
In this section, we describe how we compute the posterior distribution, $\rm P( M_{vir} | \, {\bf d})$, of the host halo mass \Mvir, given the observational data $\rm {\bf d}$.  In principle, this is obtained from the marginalisation,
\begin{equation}\label{eqn:Mvir_margpost}
\rm P( M_{vir} | \, {\bf d}) = \int P( {\bf x}, M_{vir} | \, {\bf d}) \, d {\bf x}
\end{equation}
where $\rm P( {\bf x}, M_{vir} | \, {\bf d}) = P( \bm{\theta} | \, {\bf d})$ is the joint posterior distribution of the physical parameters $\bm{\theta} =\{{\bf x}, \rm ~M_{vir}\}$ of a host-satellite system.  The parameters ${\bf x}$ are the true, latent values of the observable satellite subhalo properties, and consist of:
\begin{itemize}
\item $\rm v_{max}$, the observed maximum circular velocity of a satellite (either the LMC or M33), 
\item $\rm r$, its position relative to its host, 
\item $\rm j$, the total specific orbital angular momentum, and 
\item $\rm v_{tot}$, the satellite's total velocity relative to its host galaxy (the MW or M31).
\end{itemize}
The observable parameter vector ${\bf x}$ is a subset of these properties that depends on the type of analysis we perform (as described in Section~\ref{subsubsec:likelihoodfns}).  The observational data ${\bf d}$ consist of the measurements of the parameters in ${\bf x}$.  (If measurement errors were zero, then ${\bf d} = {\bf x}$).  The superscript $\rm^{obs}$ will indicate the observed values that remain constant in this analysis. For example, $\rm r^{obs}$ is the observed measurement of the true distance $\rm r$.  See Table~\ref{table:bayesparams}.

The joint posterior distribution is computed from the likelihood and prior via Bayes' theorem:
\begin{equation} \label{eq:bayes}
\rm P(\bm{\theta}|\,{\bf d}) \propto P({\bf d}|\,\bm{\theta}) \, P(\bm{\theta}). 
\end{equation}
The prior $\rm P(\bm{\theta})$ encodes the correlations between the observable parameters ${\bf x}$ and \Mvir, as determined by the physics of galaxy formation and evolution in the Illustris-Dark simulation.  The likelihood $\rm P({\bf d}|\,\bm{\theta})$ constrains the values of the physical parameters consistent with the measurements ${\bf d}$ of a particular subhalo (LMC or M33).  The posterior combines the prior and likelihood to obtain constraints on the \Mvir of the halo (MW or M31).  

In practice, we compute the posterior distribution (Eq. \ref{eq:bayes}) using a technique called importance sampling.   We treat halo analogs from the Illustris-Dark simulation, with physical parameters $\bm{\theta}$, as draws from the prior, which are then weighted by a likelihood function, $\rm P({\bf d}|\, \bm{\theta})$, in proportion to their similarity to the observed measurements $\rm \bf{d}$.  The resulting importance weights are used to compute posterior inferences on the virial mass, \Mvir.

In the following sections, we describe the selection criteria (denoted by ${\bf C}$) for the prior, how we calculate the appropriate likelihoods and importance weights for host haloes in Illustris-Dark, and how we compute the resulting posterior inferences for host halo mass from the observational data.

\subsubsection{Prior}
\label{subsubsec:prior}
\begin{figure*}
\begin{center}
\includegraphics[scale=0.77]{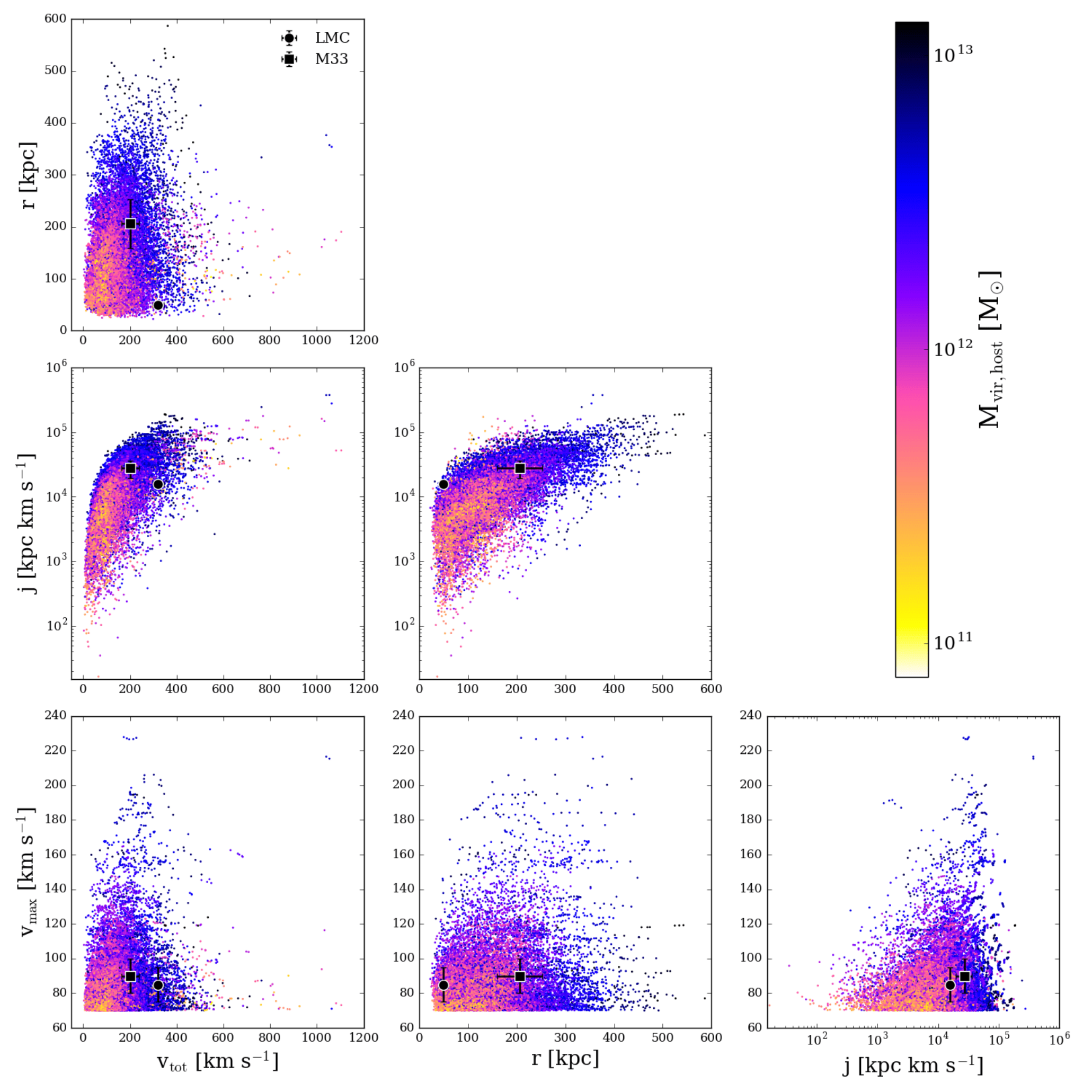}
\caption{For all host-satellite systems selected from Illustris-Dark (P($\theta$)), the distribution of the satellite subhalo properties ({\bf x}) are shown for each pair of satellite parameters. All points are colored by the corresponding host halo virial mass (\Mvir) to which they belong. The LMC's properties are indicated by a black circle, while M33 is represented with a black square. These reference points indicate that host haloes in Illusrtris-Dark do host massive satellite analogs with properties similar to that of the LMC and M33. Total orbital angular momentum ($\rm j^{obs}$) suggests that M33 should reside in a higher host halo mass than the LMC, however, similar conclusions cannot be drawn for $\rm v_{max}^{obs}$, $\rm r^{obs}$, and $\rm v_{tot}^{obs}$.
\label{fig:priorscatter}}
\end{center}
\end{figure*}

The prior $\rm P(\bm{\theta})$ is a collection of haloes from the full Illustris-Dark halo catalog. Several selection criteria (${\bf C}$) are applied to the halo catalog to choose haloes that host appropriate analogs of the LMC and M33. Those haloes that satisfy them are treated as draws from the underlying prior distribution. Therefore, the prior is truly $\rm P(\theta|\, {\bf C})$. Note that the prior is different from the control sample used in Section~\ref{sec:illustris}.

To infer the most typical host halo mass for the MW and M31 given that they both host at least one massive satellite galaxy, we must first apply some selection criteria, ${\bf C}$, to the Illustris-Dark halo catalog. ${\bf C}$ restricts the Illustris halo catalog by requiring the following criteria. 
\begin{itemize}
\item[] C$_1$: A subhalo is considered a massive satellite analog only if \indent\indent $v_{\rm max} >$ 70 \kms.
\item[] C$_2$: The massive satellite analog must reside within its host's \indent\indent virial radius (\Rvir) at $z\approx0$.
\item[] C$_3$: The massive satellite analog must have a minimal subhalo \indent\indent mass of $10^{10}$ \Msun at $z\approx0$. 
\end{itemize}
Only host-satellite systems where exactly $\rm N_{sub}$=1 massive satellite analog satisfies these qualifications are considered. All other systems (i.e $\rm N_{sub}$=0 or $\rm N_{sub}$>1\footnote{Note that G13 does weigh the consequences of including any number of subhaloes in their selection criteria for their prior sample. They found insignificant changes to the resulting MW and Local Group mass when considering exactly two MC subhalo analogs versus any number of subhalos in their analysis.}) are dismissed from the prior.

We build the prior by searching for all systems that fit these criteria over a redshift window of $ z=0-0.26$. This redshift window corresponds to 20 Illustris -Dark simulation outputs, or equivalently 60 snapshots of the Bolshoi simulation output, as B11 and G13 have used. Note that we only search for additions to the prior sample across this redshift range to increase the number of systems that could be analogs of the MW/M31 at present day. We find 19,653 systems over this redshift range that satisfy the selection criteria (${\bf C}$). Throughout the rest of this work, we use only this sample of host-satellite analogs (the prior PDF,  $\rm P({\bf \theta})$ in Eq.~\ref{eq:bayes}) to find the probability distribution of host halo mass.

There are several differences in the selection criteria for specific satellite properties in G13, B11, and in this work that should be noted. First, we alter the value of $\rm v_{\rm max}$ used for the lower bound on the prior sample. We have increased this value from 50 \kms to 70 \kms because $\sim$ 70 \kms corresponds to the maximal circular velocity for an $8 \times 10^{10}$ \Msun halo approximated with an NFW density profile. Since $8\times10^{10}$ \Msun is the lower mass bound on our massive satellite analogs sample in Paper I from abundance matching, we also use it here for consistency.

An extra mass floor, which requires each subhalo to be at least $10^{10}$ \Msun at $z=0$, is also imposed since observations show that the dynamical masses of both the LMC and M33 are greater than this value \citep[e.g.,][]{corbelli03, majewski09, saha10, corbellisalucci, vdmnk14}. 

Finally, we require each massive satellite to be within the virial radius of its host instead of within 300 kpc (G13/B11 method). Since the virial radius evolves with time, we choose this criteria instead of an arbitrarily fixed position. In all, the method described here includes as much known information about the true properties of the LMC and M33 to infer host halo mass with the simulation data while leaving the host halo mass itself (\Mvir) as a free parameter. 

The distribution of properties for all host-satellite systems in the prior can be see in Fig.~\ref{fig:priorscatter}. All pairs of observable parameters are plotted for the satellite subhaloes and each point is colored by the corresponding host halo mass from Illustris-Dark. Notice that the colorbar encompasses more than two orders of magnitude for host halo mass. The LMC and M33 are indicated by a black circle and square, respectively, on each panel. The value of $\rm j^{obs}$ for M33 seems to indicate that it should reside in a higher host halo mass than the LMC, while $\rm r^{obs}$ and $\rm v_{tot}^{obs}$ do not illustrate the same trend. Therefore, we generally expect satellites with higher total angular momenta to reside in higher mass host haloes. Our subsequent analysis allows for the combination of satellite properties to statistically infer the most likely host halo masses.

\subsubsection{Likelihood}
\label{subsubsec:likelihoodfns}

In Eq.~\ref{eq:bayes}, $\rm P({\bf d}| \, \bm{\theta})$ is the sampling distribution of the measured data {\bf d} given the physical parameters $\bm{\theta} = \{{\bf x}, \rm ~M_{vir}\}$.  However, this only depends on the true values of the observables ${\bf x}$, and the measurement error distribution.  Equivalently, given ${\bf x}$, the data ${\bf d}$ is conditionally independent from \Mvir.  The individual satellite properties for the LMC and M33 are treated as independent measurements as the covariance between the observed position and velocity of a given satellite was shown to be significantly smaller than the variances on the measurements in B11. Therefore, $\rm P({\bf d}| \, \bm{\theta}) = \rm P({\bf d}| \, {\bf x})$.   When viewed as a function of the parameters with the observed data fixed, this factor is the joint likelihood function, $\mathcal{L}({\bm{\theta} | \, {\bf d}}) = \mathcal{L}({\bf x} | \, {\bf d})$. 
$\mathcal{L}({\bf x| \,d})$ is simply a product over the individual data, $\rm d_i$:
 \begin{equation} 
 \rm  \mathcal{L}({\bf x|\, d}) = P(\textbf{d}|\,\textbf{x})= \prod_{i}^m P(d_i |\,x_i), \label{eq:L1}
 \end{equation}
We construct two different likelihoods that each utilise a different set of satellite properties. One main difference between the likelihood function in G13/B11 and this paper is that they build a joint likelihood based on the existence of two massive satellites (analogs of the MCs) and their subsequent observed properties (such that $m=6$ properties), whereas we only require host haloes to have one massive satellite (and thus $m=3$ properties for the instantaneous method and $m=2$ for the momentum method). We do not include more than one massive satellite in this analysis because we generalize this method for application to both the MW and M31. 

Furthermore, for the MW's mass estimate, if we require all prior haloes to contain both an LMC and SMC analog, the sample size effectively reduces to approximately zero. As we discuss in Section~\ref{subsec:observederrors}, the rarity of the LMC's (and SMC's) current orbital configuration alone reduces the number of haloes in the prior that contribute to the inference scheme. Additionally, simply requiring host haloes to contain MC analogs based on the three observed properties discussed above does not account for the binarity and shared orbital trajectories of the MCs, so we omit these criteria and the SMC from this analysis.
 
\vspace{.2cm}
\noindent  { \em I. Instantaneous Likelihood}
\vspace{.2cm}

The Instantaneous Likelihood uses as the observable parameters {\bf x}: the satellite's maximum circular velocity $\rm v_{max}$, its separation $\rm r$ from the host, and the total velocity today relative to the host galaxy $\rm v_{tot}$.  The data {\bf d} are the observed measurements of these quantities (Table~\ref{table:bayesparams}).

 \begin{equation}\rm \mathcal{L}({\bf x|\, d})= N(v_{max}^{obs} |\, v_{max}, \sigma_v^2) \times N(r^{obs} |\, r, \sigma_r^2) \times N(v^{obs} |\, v_{tot}, \sigma_v^2), \label{eq:L2} 
 \end{equation}
where 
\begin{equation} \rm N(y |\, \mu, \sigma) = \frac{1}{\sqrt{2\pi \sigma^2}} exp\left[\frac{-(y-\mu)^2}{2\sigma^2}\right]. \end{equation}
is a Gaussian probability density for random variable $\rm y$ with mean $\mu$ and variance $\sigma^2$.
The $\sigma$ quantities are the standard deviations of the measurement errors of the corresponding observations. We use this likelihood to compare with the results of G13 and B11.

\vspace{.2cm}
\noindent { \em II. Momentum Likelihood }
\vspace{.2cm}

 Our second method for computing the joint likelihood uses a different subset of the satellite parameters {\bf x} and data ${\bf d}$, focusing more on orbital dynamics.  Our Momentum Likelihood is based on only two parameters: the satellite's $\rm v_{\rm max}$ and the magnitude of its orbital angular momentum, $\rm j$. Fig.~\ref{fig:priorscatter} demonstrates that these properties are also only very weakly covariant, so they can be approximated as independent measurements. The angular momentum likelihood is therefore,
\begin{equation}\rm \mathcal{L}({\bf x|\, d}) = N(v_{max}^{obs} |\, v_{max}, \sigma_v^2) \times N(~j^{obs} |\,j, \sigma_j^2), \label{eq:L3} \end{equation}
where $\rm j$ is the magnitude of the orbital angular momentum vector. The mean and the standard deviation on $\rm j$ (i.e. $\rm j^{obs}$ and $\rm\sigma_j$) for the LMC and M33 are computed from the 10,000 Monte Carlo samples described in Section~\ref{subsec:observed}.

Since orbital angular momentum is generally stable over time compared to other orbital parameters, we investigate how closely this likelihood construction agrees with the B11 method.  We still use the same draws from the Illustris-Dark halo catalog as described in Section~\ref{subsubsec:prior} as the prior, but we change how the importance sampling weights are computed from the likelihood, as described below.

\subsubsection{Importance Sampling}

Now that the prior and likelihood have been defined, we return to Bayes' theorem
\begin{equation}\label{eq:bayes4} 
\rm P({\bf x}, M_{vir}|\,{ \bf d, C}) \propto P({\bf d}| \, {\bf x}) \times P({\bf x}, M_{vir}|\, {\bf C}), 
\end{equation}
where we explicitly denote the dependence on the prior selection criteria ${\bf C}$.

With the prior and likelihoods defined, the marginal distribution of \Mvir and therefore the posterior distribution for the halo mass of the MW and M31 can now be computed using this form of Bayes' theorem.

The posterior PDF is computed using a technique called importance sampling. In importance sampling, a set of samples is drawn from an importance sampling function and weighted accordingly while calculating integrals over the posterior PDF (see B11 and references therein). The importance sampling function is chosen to be the prior PDF, as in B11, so that our weights are proportional to the likelihoods. Using these importance weights, we can calculate integrals summarising the posterior PDF for our target parameter -- the host galaxy's halo mass. These integrals describe the mean halo mass, credible intervals surrounding the mean, and a representation of the  marginal posterior PDF for host halo mass, Eq. \ref{eqn:Mvir_margpost} (in the form of counts per  $\rm dM_{vir}$).

Expectations of functions of the physical parameters under the posterior PDF are approximated as sums over the $n$ samples as follows:
 \begin{equation}
 \begin{split}
 \rm  \int f(\bm{\theta}) P({\bf x}, M_{vir}|\, {\bf d, C}) \, d\bm{\theta} &= \frac{\int \rm f(\bm{\theta})P({\bf d}|\,{\bf x})P({\bf x}, M_{vir}| {\bf C}) \, d\bm{\theta}}{\int \rm P({\bf d}|\,{\bf x}) P({\bf x}, M_{vir}|\, {\bf C}) \, d\bm{\theta}} \\
&\approx \frac{\sum_j^n \rm f(\bm{\theta}_j)P({\bf d}|\,{\bf x}_j)}{\sum^n_j \rm P({\bf d}|\,{\bf x}_j)}. 
\end{split}
\end{equation}
The denominator of this equation is the normalization constant. If the chosen $f(\theta)$ depends only on \Mvir, then the final sum implicitly computes an expectation with respect to the marginal posterior of \Mvir:
\begin{equation} \label{eq:PDF}
\begin{split}
\rm  \int \rm f(M_{vir}) P(M_{vir}|\, {\bf d, C}) \, dM_{vir} \\ 
= &\int \rm f(M_{vir})\, \rm P({\bf x}, M_{vir}|\, {\bf d, C}) \, d{\bf x} \, dM_{vir} \\
\approx & \frac{\sum_j^n \rm f(M_{vir}^j) \, P({\bf d}| \,{\bf x}_j)}{\sum^n_j \rm P({\bf d}|\,{\bf x}_j)} \\
=&\sum_j^n \rm f(M_{vir}^j) \, w_j \\
\end{split}
\end{equation}
where $\rm w_i = P({\bf d}|\,{\bf x}_i)/\sum^n_j \rm P({\bf d}|\,{\bf x}_j)$ are importance weights. The weights derived from the likelihood function represent the degree to which subhalo properties in Illustris-Dark resemble the observed properties of the LMC and M33 and consequently how much each halo contributes to the posterior probability density function (PDF) for the halo mass of the MW or M31.

Setting $f(\bm{\theta}) =$ \Mvir 
gives the posterior mean value for host halo mass of the MW or M31. To create a representation over the full posterior PDF, Eq.~\ref{eq:PDF} is computed for contiguous intervals in host halo mass. For example, to calculate the posterior probability that the host halo mass is between $1-1.5 \times 10^{12}$ \Msun, set $f({\bf x}) = 1$ for all Illustris-Dark haloes in the prior that satisfy this fiducial range or let $f({\bf x}) = 0$ otherwise. Repeating this method for many fiducial halo mass ranges results in a coarsely sampled representation of the posterior PDF in a histogram-like fashion. For more details on how we create a smooth representation of the posterior PDFs, see Appendix~\ref{sec:appendixB}, which describes the kernel density estimation technique used here.

In practice, it is more convenient to compute and report summaries on a log scale, i.e. $\rm P( \log_{10} M_{vir} | {\bf x}, {\bf C})$\footnote{ $\rm \log_{10} M_{vir}$ should be interpreted as  $\rm\log_{10} (M_{vir}$/\Msun).} rather than $\rm P( M_{vir} | {\bf x, C})$. This is because the former is more roughly Gaussian, and thus more easily summarised by a central value and width, whereas the latter is non-Gaussian with a skewed right tail.  Hence, we summarise the posterior PDF of $\rm \log_{10} M_{vir}$ with its posterior mean and 68 per cent highest posterior density credible intervals. When we report the mass estimates on a physical scale as $\rm M_{vir} = X^{+U}_{-L}$ \Msun, these should be interpreted on a log scale, such that $\rm log_{10} X$ is the posterior mean of $\rm \log_{10} M_{vir}$, and [$\rm \log_{10}(X-L), \log_{10}(X+U)$] is the 68 per cent credible interval in $\rm log_{10} M_{vir}$ with the highest posterior density\footnote{We caution that these summaries on the log scale should not be naively translated to constraints on the linear scale.  For example, the posterior mean of $\rm\log_{10} M_{vir}$ is generally not equivalent to the $\rm log_{10}$ of the posterior mean of $\rm M_{vir}$, as probability densities do not trivially transform under a nonlinear change of variables.}.

\section{MW and M31 Mass Results From Massive Satellite Properties}
\label{sec:results}
Following the statistical method described in Section~\ref{sec:bayesian}, we present the posterior distributions for the halo mass of the MW and M31 based on the dynamics of their most massive satellites. The posterior distributions have been computed for two different likelihood functions (instantaneous vs. momentum; Section~\ref{subsubsec:likelihoodfns}). We also examine the robustness of the two methods as a function of time and satellite orbital history.

\begin{figure*}
\begin{center}
\includegraphics[scale=0.7]{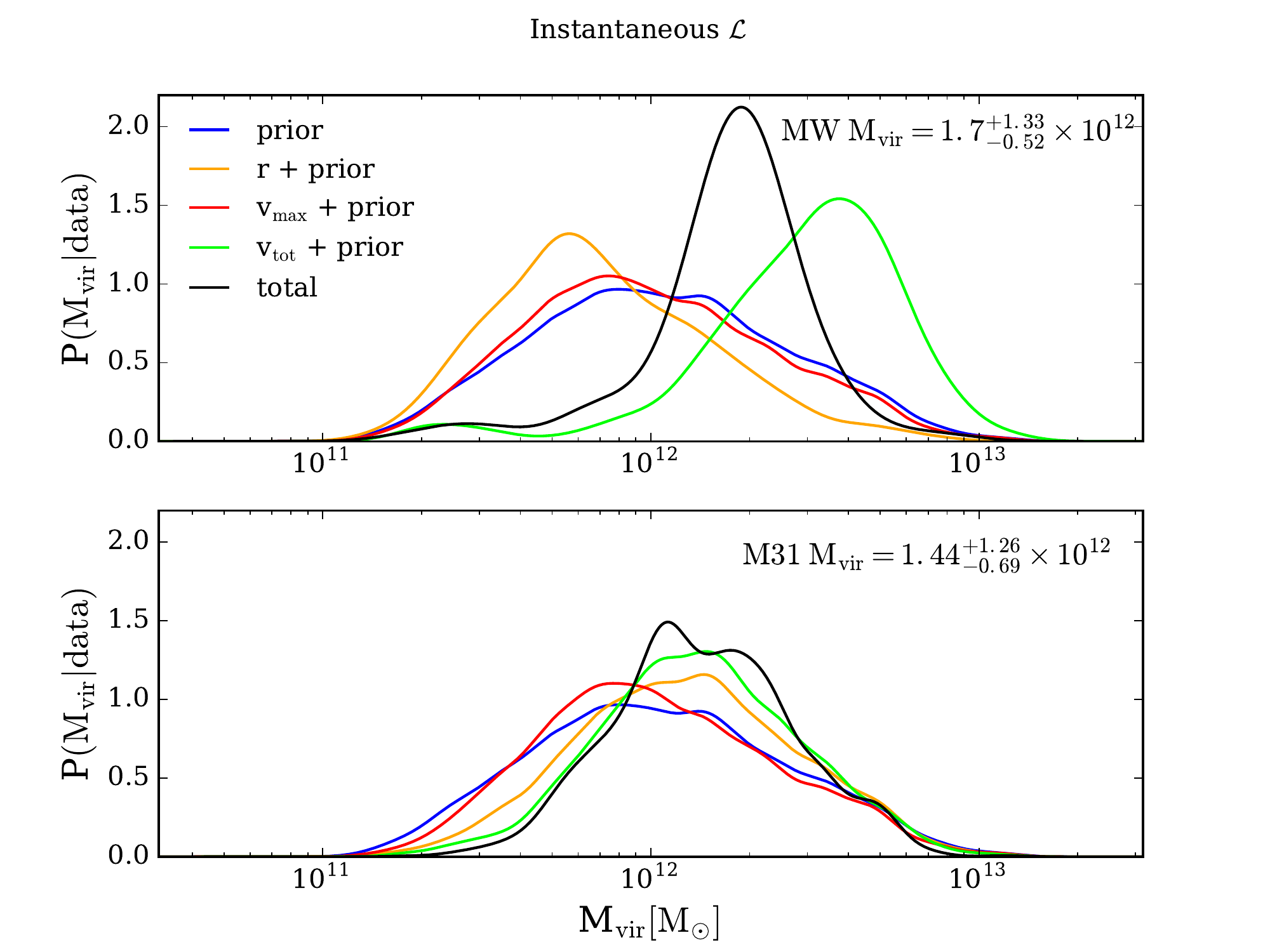}
\caption{The posterior distribution of the MW (top) and M31's (bottom) halo mass inferred from the properties of their brightest and most massive satellites, the LMC and M33. The solid lines show the posterior PDFs calculated with the Illustris-Dark halo catalog based on the following properties: (a) the existence of exactly one satellite with $\rm v_{max} >$ 70 \kms, a $z=0$ dark matter mass $\geq 10^{10}$ \Msun, and residing within the virial radius of its host (blue), (b) the maximum circular velocities $\rm v_{max}$ of the LMC or M33 (red), (c) the distance of the satellite from the centre of its host (orange), (d) the velocity of the satellite relative to the host (green), and (e) all of these properties. The set of host-satellite haloes drawn from the Illustris-Dark halo catalog passing the selection criteria (${\bf C}$) give a combined (black solid line) MW halo mass $\rm M_{vir} = 1.70^{+1.33}_{-0.52} \times 10^{12}$ \Msun, or $\rm \log_{10}M_{vir}=12.23^{+0.25}_{-0.16}$ (68 per cent credible interval). Using the instantaneous position and velocity of M33, the halo mass of M31 is $\rm M_{vir}=1.44^{+1.26}_{-0.69} \times 10^{12}$ \Msun, or $\rm \log_{10}M_{vir}=12.16^{+0.27}_{-0.28}$ (68 per cent credible interval).
\label{fig:instbayes}}
\end{center}
\end{figure*}

\begin{figure*}
\begin{center}
\includegraphics[scale=0.7]{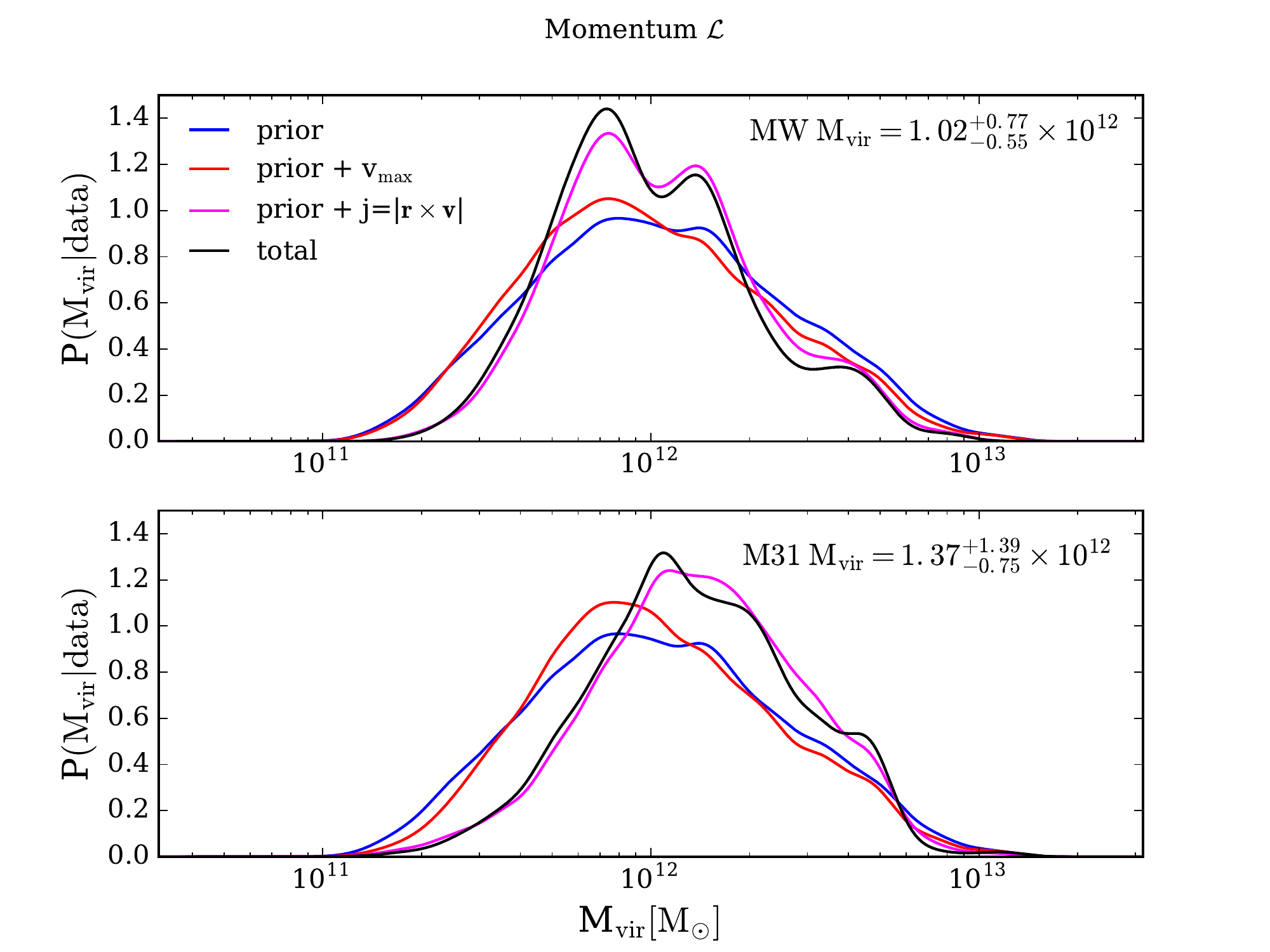}
\caption{The posterior PDFs for the inferred halo mass of the MW (top) and M31(bottom) based on (a) the existence of exactly one satellite with $\rm v_{max} >$ 70 \kms, a $z=0$ dark matter mass of $\geq 10^{10}$ \Msun, and residing within the virial radius of its host (blue), (b) the maximum circular velocities $\rm v_{max}$ of the LMC or M33 (red), (c) the magnitude of orbital angular momentum (magenta) for the satelltie, and (d) all of these properties combined (black). The total posterior PDF infers a MW halo mass $\rm M_{vir} = 1.02^{+0.77}_{-0.55} \times 10^{12}$ \Msun, or $\rm \log_{10}M_{vir}=12.01^{+0.25}_{-0.34}$ (68 per cent credible interval). Using the orbital angular momentum of M33, the halo mass of M31 is $\rm M_{vir}=1.37^{+1.39}_{-0.75} \times 10^{12}$ (bottom), or $\rm \log_{10}M_{vir}=12.12^{+0.32}_{-0.35}$ (68 per cent credible interval). In contrast to the results of Fig.~\ref{fig:instbayes}, here hosts of M33 satellites are found to be more massive than those that host satellites with properties similar to the LMC. In general, the momentum method also results in broader constraints on the mass of the MW and M31, respectively.
\label{fig:mombayes}}
\end{center}
\end{figure*}

\subsection{Bayesian Inference with Instantaneous Satellite Kinematics}
\label{subsec:kinematics}
From the observed data in Table~\ref{table:bayesparams} ($\rm v_{max}^{obs}, r^{obs},  v_{tot}^{obs}, j^{obs}$) and the statistical method described above, we find posterior mean values for the most likely halo mass of the MW and M31. Fig.~\ref{fig:instbayes} shows the posterior distribution of the resulting MW and M31 halo masses using the observed properties of the LMC and M33, respectively, as inputs to the instantaneous likelihood function (Eq.~\ref{eq:L2}). The individual curves represent the posterior PDFs based on specific satellite parameters. For example, posteriors are calculated based on the full prior sample (blue), $\rm v_{max}$ (red), $\rm r$ (orange), $\rm v_{tot}$  (green), and all satellite properties combined (black). 

For the host haloes weighted for subhalo properties most like the LMC, we find that the preferred halo mass for the MW is $\rm M_{vir} = 1.70^{+1.33}_{-0.52} \times 10^{12}$ \Msun (top panel of Fig.~\ref{fig:instbayes}). Applying the same rationale using the properties of M33, we find the most typical host halo mass for M31 is $\rm M_{vir} = 1.44^{+1.26}_{-0.69} \times 10^{12}$ \Msun (bottom panel of Fig.~\ref{fig:instbayes}). Interestingly, the inferred mean value for the MW is higher than that of M31. This is likely due to the high relative velocity of the LMC today compared to that of M33, thereby suggesting that the instantaneous method is not reliable. The link between satellite orbital phase and the resulting host halo mass estimates will be discussed in further detail in Section~\ref{subsec:orbits}.

Comparing the two host-satellite systems, we find that the inferred halo mass is highly correlated to the uniqueness of the combined observed satellite parameters (Table~\ref{table:bayesparams}). Since the LMC has a small relative separation and a high velocity relative to the MW at present day, the total posterior (black solid line) for the MW's halo mass is approximately centred between the contributions from the position (orange solid line)  and velocity (green solid line) posteriors. M33's present-day position and velocity are much more typical in a population of massive satellites (see top left panel of Fig.~\ref{fig:priorscatter}), so we find that the individual and total posteriors are all in good agreement with each other. 

While our prior is composed of $\sim$20,000 haloes, it is important to know how many of these haloes actually contribute to this statistical inference. Table~\ref{table:ess} indicates how many haloes in the prior host subhaloes with properties within an average of $1\sigma$ and $2\sigma$ on the observed properties of the LMC and M33, respectively. The final column shows the effective sample size (ESS) of each likelihood method for each of the satellites. The ESS is the number of haloes that actually statistically contribute to the importance sampling and therefore most heavily influence the posterior PDF of host halo mass. See~Appendix~\ref{sec:appendixB} for more details.

As the instantaneous properties of the LMC today ($\rm v_{max}, r,  v_{tot}$) are rare, very few haloes in the Illustris-Dark prior host LMC analogs that exhibit this specific combination of observed satellite properties. Consequently, the ESS is low and few haloes statistically determine the posterior halo mass of the MW for the instantaneous likelihood method. To determine the sampling noise on the inferred MW mass with the instantaneous likelihood method, we create 25 bootstrap resampled mock catalogs from the original prior described in Section~\ref{subsubsec:prior} and recompute the MW's mass using the instantaneous likelihood method. By doing so, we can separate how much additional uncertainty on the posterior mean mass of the MW comes from the Monte Carlo error caused by a small ESS. The standard deviation of the posterior mean MW mass from the 25 mock catalogs is $0.13\times10^{12}$ \Msun. For the instantaneous likelihood method using M33's observed properties, we find that there is minimal ($0.01\times10^{12}$ \Msun) additional uncertainty associated with the posterior mean mass of M31 due to a small ESS. The ESS is significantly high as M33's observed instantaneous properties are not rare like the LMC's.

\begin{table}
\centering
\caption{The number of haloes in the prior that contribute to the statistical inference. A total of 19,563 haloes are considered for the analysis. The second column lists the number of haloes with massive satellites within an average of 1$\sigma$ on the subset of observed properties for the LMC and M33 used in the instantaneous ($\rm v_{max}, r,  v_{tot}$) and momentum ($\rm v_{max}, j$) likelihood methods. The third column provides the number of haloes with satellites exhibiting properties within the 2$\sigma$ range of the observed values used for both likelihoods. The final column provides the effective sample size (ESS, see Eq.~\ref{eq:ess}), which is the number of statistically relevant haloes for importance sampling.}
\label{table:ess}
\begin{tabular}{l|c|c|c}
& $1\sigma$ & $2\sigma$ & ESS \\ \hline
MW/LMC Instantaneous & 10 & 56 & 42 \\ \hline
M31/M33 Instantaneous & 503 & 5,902 & 3,033 \\ \hline \hline
MW/LMC Momentum & 971 & 3,459 &3,465 \\ \hline
M31/M33 Momentum  & 1,347 & 8,017 & 8,143 \\ \hline
\end{tabular}
\end{table}

\subsubsection{Implications for Different M31 Proper Motion Measurements and the Instantaneous Method}
In Paper I, we explored the implications for different values of M31's proper motion component on the orbital history of M33. We use the M31 proper motion reported by vdM12 throughout Paper I and this work. The vdM12 results are an extension of the direct measurement of M31's proper motion with HST by S12 such that vdM12's measurement is a weighted average of the proper motion inferred from the kinematics of M31 satellites and the S12 direct measurements \citep[see also][]{vdmg08}. They find a combined proper motion of  $\rm v_{tan}=17\pm17$ \kms.

Other teams have also measured the proper motion of M31 using different techniques. \citet[][hereafter S16]{salomon16} recently inferred the tangential proper motion of M31 using the motions of its satellites and find a value of $\rm v_{tan}\sim150$ \kms. Here, we compute the posterior distributions for M31's halo mass where $\rm v_{tot}^{obs}$ and $\rm\sigma_v$ from Table~\ref{table:bayesparams} are changed to reflect the total velocity of M33 relative to M31 using the M31 proper motion measured by each of S12 and S16 independently. 

Using only the S12 M31 proper motion, rather than the average vdM12 reported value, M33's velocity relative to M31 becomes $\rm v_{tot}^{obs}=242$ \kms and $\rm\sigma_v=76$ \kms. The mean M31 halo mass inferred using these velocity values is \Mvir$=1.48^{+1.70}_{-0.77}\times10^{12}$ \Msun. As expected, the mean value and 68 per cent credible interval of M31's halo mass increases compared to the top panel of Fig.~\ref{fig:instbayes} since $\rm v_{tot}^{obs}$ and $\rm\sigma_v$ both increase. 

For the S16 tangential velocity, $\rm v_{tot}^{obs}=139$ \kms and $\rm\sigma_v=52$ \kms, which is substantially lower than both the value listed in Table 1 and the S12 results. Using these values as inputs to the instantaneous likelihood function (Eq.~\ref{eq:L2}), we find that M31 \Mvir$=1.03^{+0.82}_{-0.55}\times 10^{12}$ \Msun. The decrease in total velocity reduces the posterior mean mass for M31 significantly and the 68 per cent credible interval also shifts towards lower values. Overall, both of the mean values resulting from the two different tangential velocity measurements are encompassed within the 68 per cent credible interval of the original posterior mean mass of M31 determined with the properties listed in Table~\ref{table:bayesparams} and the instantaneous likelihood method. However, these results seem to favor the S12 HST proper motions over the S16 results.

\subsection{Bayesian Inference with Angular Momentum}
\label{subsec:angmom}
We have now replaced the posterior distributions in instantaneous position and velocity by a single posterior describing the orbital angular momentum (Section~\ref{subsubsec:likelihoodfns}). The posterior distributions resulting from the orbital angular momentum likelihood function are shown in Fig.~\ref{fig:mombayes}.  The posterior distributions are shown based on the prior (blue), $\rm v_{max}$ (red), j (magenta), and all of those properties combined (black). By weighting the host haloes based on the LMC's properties, we find that the most typical halo mass for the MW is $\rm M_{vir} = 1.02^{+0.77}_{-0.55} \times 10^{12}$ \Msun (top panel). Weighting the host haloes by M33's properties, we find the most typical halo mass for M31 is $\rm M_{vir} = 1.37^{+1.39}_{-0.75} \times 10^{12}$ \Msun (bottom panel). 

In this likelihood construction, the halo masses for the MW and M31 are as expected, with M31 being more massive (see Section~\ref{sec:intro}). Overall, the two methods agree in that the inferred host halo masses and errors still encompass the same broad range of mass from the literature. For the MW's mass, the combination of position and velocity versus angular momentum causes more drastic differences in the posterior $mean$ values for halo mass compared to the results for M31. This disparity is likely due to the short-lived current position and velocity of the LMC versus its orbital angular momentum, which is fairly common amongst massive satellite analogs. We will further explore the change in inferred MW mass as a function of the LMC's orbital history in Section~\ref{subsec:bayesianevolution}.

Unlike the instantaneous method, we find that there is insignificant additional uncertainty on the posterior mean mass of the MW and M31 with the momentum likelihood method due to Monte Carlo error ($\sim$$0.01\times10^{12}$ \Msun). For both host-satellite systems, the ESS is significantly high (see Table~\ref{table:ess}) and the 25 bootstrap resampled mock catalogs provide results that are in very good agreement with those from the original prior.

\subsubsection{Implications for Different M31 Proper Motion Measurements and the Momentum Method}

We now repeat our momentum method calculations for the mass of M31 using the tangential velocities reported by S12 and S16, respectively. Using the S12 $\rm v_{tan}$ value, we have repropagated the errors in distance, radial velocity, and proper motion to calculate a total orbital angular momentum value of $\rm j^{obs} =28,940$ kpc \kms with $\sigma=10,062$ kpc \kms. With this $\rm j^{obs}$ value and its associated error, the inferred mass of M31 is \Mvir=$1.33^{+1.44}_{-0.74}\times10^{12}$\Msun. This result is in very good agreement with the vdM12 results listed above, as expected, since the vdM12 value is derived from S12.

For the S16 $\rm v_{tan}$ value, the average of the total observed angular momentum is $\rm j^{obs} =28,278$ kpc \kms with $\sigma=3,739$ kpc \kms. These properties yield an M31 mass of \Mvir=$1.65^{+1.58}_{-0.84}\times10^{12}$ \Msun. This is the highest M31 mass inferred thus far in this work, and it does not strictly conform to the trend we observed with M33 and the LMC earlier in Section~\ref{subsubsec:prior}, where the satellite with higher orbital angular momentum suggests a higher host halo mass. The S12 $\rm v_{tan}$ value provides the highest total orbital angular momentum for M33 (though only by a few percent) but it does not result in the highest corresponding M31 mass. A more precise direct measurement of M31's proper motion will better constrain $\rm v_{tan}$ and therefore $\rm j^{obs}$, thereby providing more precise measurements of M31's mass in a statistical fashion.

A summary of all posterior mean halo masses included in the 68 and 90 per cent credible intervals for the MW and M31 are presented in Table~\ref{table:results}.

\begin{table*}
\caption{The values for posterior halo mass of the MW and M31 included at the 68 and 90 per cent credible intervals for all likelihood functions explored in this analysis.}
\label{table:results}
\centering
\begin{tabular}{lcc|cc} \hline \hline
& {\bf MW \Mvir [$\rm 10^{12}\; M_{\odot}$]} & & {\bf M31 \Mvir [$\rm 10^{12}\;M_{\odot}$] }& \\ \hline \hline
& 68 per cent & 90 per cent & 68 per cent & 90 per cent \\ \hline
Instantaneous $\mathcal{L}$ & $1.70^{+1.33}_{-0.52}$ & $1.70^{+2.89}_{-1.07}$ & $1.44^{+1.26}_{-0.69}$ &  $1.44^{+2.74}_{-0.95} $ \\ \hline
\citet{sohn12} M31 $\rm v_{tot}$ & -- & -- & $1.48^{+1.70}_{-0.77}$& $1.48^{+3.55}_{-1.0}$ \\ \hline
\citet{salomon16} M31 $\rm v_{tot}$ & -- & -- & $1.03^{+0.82}_{-0.55}$ & $1.03^{+2.03}_{-0.71}$\\ \hline \hline
Momentum $\mathcal{L}$ & $1.02^{+0.77}_{-0.55} $ & $1.02^{+2.4}_{-0.70} $ & $1.37^{+1.39}_{-0.75}$ & $1.37^{+3.67}_{-0.91} $ \\ \hline
\citet{sohn12} M31 $\rm v_{tot}$ & -- & -- & $1.33^{+1.44}_{-0.74}$& $1.33^{+3.74}_{-0.9}$ \\ \hline
\citet{salomon16} M31 $\rm v_{tot}$ & -- & -- & $1.65^{+1.58}_{-0.84}$ & $1.65^{+3.83}_{-1.04}$\\ \hline \hline
\end{tabular}
\end{table*}

\subsection{The Bayesian Inference Technique as a Function of Time}
\label{subsec:bayesianevolution}

\begin{figure*}
\begin{center}
\includegraphics[scale=0.5]{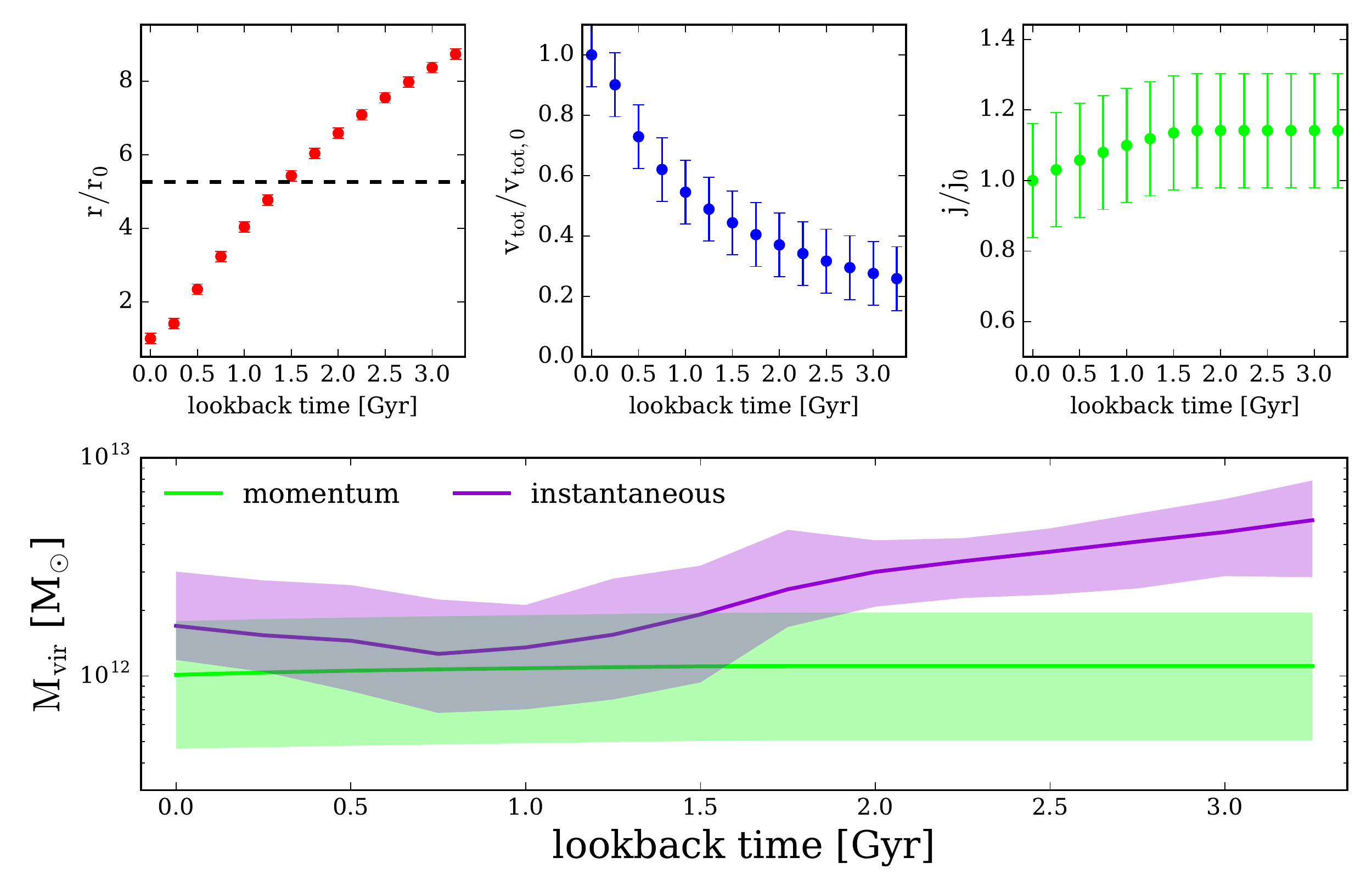}
\caption{Posterior mean mass estimates for the MW based on the orbital history of the LMC using the two likelihood functions. The top left panel shows the average past orbital history for the LMC (as calculated in Paper I) when the MW's mass is held fixed at $10^{12}$ \Msun with a virial radius of 261 kpc. This average orbital history encompasses an LMC mass range of $3-25\times10^{10}$ \Msun. The relative distance between the LMC and the MW is shown as a ratio with its current distance. The virial radius of the MW in the LMC orbital model is shown by the black dashed line. The top middle panel shows the velocity of the LMC along its orbit relative to its $z=0$ velocity, while the top right-most panel shows the total orbital angular momentum of the LMC along its trajectory relative to its $z=0$ value. The error bars for the panels in the top row are propagated to reflect HST's precision on the position and velocity of the LMC today. The bottom panel shows the resulting predictions for the mass of the MW using the instantaneous (purple) likelihood and the momentum (green) likelihood as a function of orbital configuration. When the LMC's orbit is outside of the virial radius (> 1.5 lookback Gyr), the mass of the MW is naturally biased towards higher values. However, there is still a factor of two deviation in the results from the instantaneous method in just the last $\sim$ 1 Gyr, which is of order the current uncertainty in the MW's mass.}
\label{fig:loMW}
\end{center}
\end{figure*}

\begin{figure*}
\begin{center}
\includegraphics[scale=0.5]{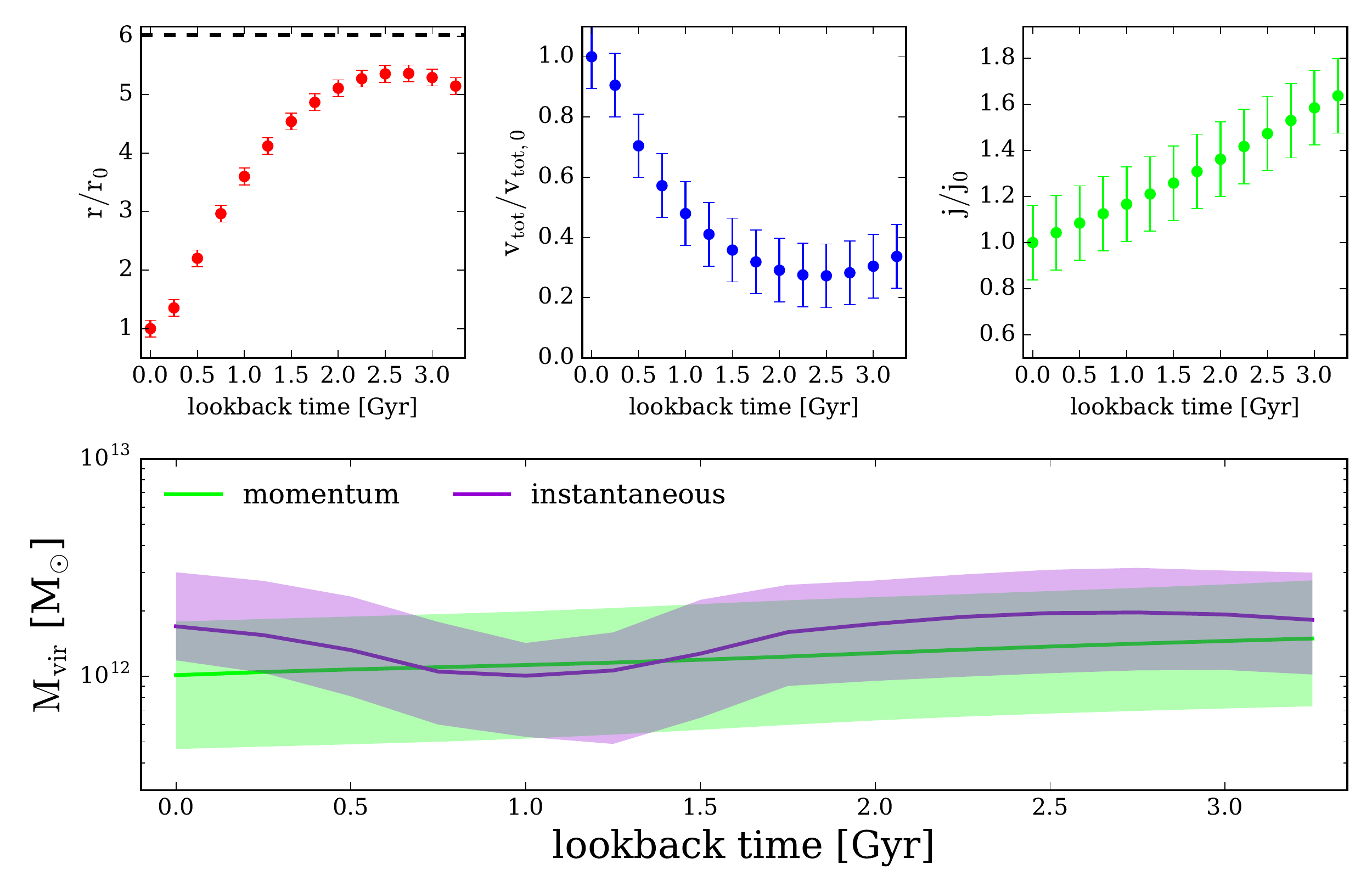}
\caption{The posterior mean mass estimates for the MW are shown using the LMC's orbital history calculated with a fixed MW mass of $1.5\times10^{12}$ \Msun and a viral radius of 299 kpc. See Fig.~\ref{fig:loMW} for more details. When the LMC is not on a first infall orbital trajectory, the instantaneous and momentum methods are in better agreement over time. However, there is still a factor of two deviation in the inferred mass of the MW with the instantaneous method, proving that it is highly sensitive to satellite orbital configuration.}
\label{fig:hiMW}
\end{center}
\end{figure*}

Thus far, the posterior distributions for the halo mass of the MW and M31 resulting from both likelihood functions have been calculated using only the observed properties of the LMC and M33 today. Orbital timescales of massive satellites are typically about 5-6 Gyr (see Paper I for orbits of the LMC and M33) and satellites will exhibit a range of positions and velocities during a single orbital period. Some satellites experience more variation than others depending on their host environment, eccentricity, and impact parameter at infall. We have already shown that satellite orbital angular momentum remains fairly well-conserved during that time in Section~\ref{subsubsec:dynamics}, aside from some angular momentum loss between infall and today due to dynamical friction. In this section, we test the robustness of the instantaneous and momentum likelihood methods as a function of time using the numerically integrated orbital histories of the LMC from Paper I.

We consider two orbital histories for the LMC -- one in a low mass MW halo ($10^{12}$ \Msun) and one in a high mass MW halo ($1.5\times10^{12}$ \Msun). Both orbital histories represent an average of orbits computed using an LMC mass range of $3-25\times10^{10}$ \Msun and the mean position and velocity of the LMC. In the low mass MW halo, the LMC is on first infall into the MW's halo.  In the higher mass halo, it has achieved a pericentre about 5 Gyr ago and remains within the virial radius of the low mass MW model (261 kpc) for the entirety of the last 6 Gyr. Full details for these orbital models can be found in Paper I.

In intervals of 0.25 lookback Gyr, we recompute the posterior distribution for the MW's halo mass using both the instantaneous and momentum methods with the properties of the LMC at each time interval along its integrated orbital trajectory. The calculations for $t=0$ lookback Gyr were computed with the LMC's properties listed in Table~\ref{table:bayesparams} and were already summarized in Section~\ref{sec:results}. This process is repeated until 3.25 Gyr ago using the position and velocity (instantaneous likelihood) or orbital angular momentum (momentum likelihood) of the LMC. We terminate the analysis at $\sim$3 Gyr ago because the host-satellite systems that constitute the prior have been chosen from a redshift range of $z=0-0.26$ ($\sim$3 Gyr).

The evolution of the statistically inferred MW halo mass for the two different LMC orbital histories are shown in Figs.~\ref{fig:loMW} and \ref{fig:hiMW}. The top left panel shows the evolution of the LMC's position relative to the MW as function of time, scaled to the LMC's position today ($\rm r_0^{obs}=50$ kpc). The top middle panel shows the velocity of the LMC along its orbit relative to its $z=0$ velocity ($\rm v_0^{obs}=321$ \kms), while the top right panel shows the evolution of its total orbital angular momentum scaled to its value today ($\rm j_0^{obs}=27,656$ kpc \kms). Given these {\em observed} LMC properties at each interval in lookback time, the posterior mean MW halo mass included in the 68 per cent credible intervals are plotted in the bottom panels of Figs.~\ref{fig:loMW} and~\ref{fig:hiMW} for the instantaneous (purple) and momentum likelihoods (green). The measurement errors for all quantities in the top panel are assigned to match the precision of the observed LMC properties today. 

Fig.~\ref{fig:loMW} shows that the posterior mean MW halo mass inferred by the instantaneous likelihood construction and a first infall scenario changes drastically as a function of time. In just 3 Gyr, the mean inferred MW mass varies from about $10^{12}$ at minimum to $\sim4\times10^{12}$ \Msun at maximum. On the other hand, the posterior mean mass of the MW remains mostly constant at 1-1.1$\times10^{12}$ \Msun when computed using the orbital angular momentum of the LMC as a function of time. The contrast between these two results clearly demonstrates how strongly the inferred mass of the MW can be biased by the satellite parameters, especially when those parameters change significantly with time. Therefore, while the LMC's position and velocity today yield a reasonable mass estimate for the MW that is in agreement with the MW mass inferred by the LMC's orbital angular momentum, this result largely hinges on the LMC's orbital phase at any given time and therefore its past orbital history. 

Note that for the first infall scenario (Fig.~\ref{fig:loMW}), the LMC does not remain within the virial radius of the adopted MW mass model for all 3 lookback Gyr. However, the satellites in the prior from which the MW mass results are calculated are all chosen such that they reside within the virial radius of their host. Therefore, in the first infall scenario, MW mass estimates determined at times when the LMC is outside of the virial radius (> 1.5 lookback Gyr) are necessarily biased to high MW mass and thus deviate the most strongly. However, the inferred mass of the MW with the instantaneous method still varies by a factor of almost two in just the last $\sim 1$ Gyr, which is of order the current uncertainty on the mass of the MW. 

We have also checked that the ESS values for each time interval are sufficiently large (a few factors greater than the ESS at $t=0$ lookback Gyr). As the position and velocity of the LMC become more common amongst the phase space of the subhaloes in the prior, the ESS increases beyond $t=0$ lookback Gyr. Therefore, the reported MW mass values included in the 68 per cent credible interval are statistically representative of the prior and not just an artifact of a low ESS value.

When the posterior mean masses of the MW are computed using a less energetic orbital history where the LMC has made a pericentric passage 5 Gyr ago, we find that the results are less dependent on any specific likelihood construction. Fig~\ref{fig:hiMW} demonstrates that the inferred posterior mean mass of the MW calculated with the instantaneous and momentum likelihoods are in much better agreement over time. However, the posterior mean mass of the MW inferred with the instantaneous method still varies by approximately a factor of two, whereas the momentum method provides consistent results with time. The 68 per cent curves for both constructions agree for a majority of the last 3 Gyr, demonstrating that the satellite's orbital trajectory is key to the robustness of these Bayesian techniques. We conclude that the instantaneous method is therefore less reliable for inferring host halo mass, regardless of the satellite's orbital energy, as was demonstrated by the two different LMC scenarios. Orbital `constants' like orbital angular momentum prove to be more reliable with time, and thus the momentum method is preferred. 

Satellites on high speed orbits will be most affected by this issue, though this may also hold true for satellites in lower energy orbits that are fairly eccentric. For example, Leo I, a MW dSph satellite that resides at a distance of about 260 kpc today, also appears to be on its first infall into the halo of the MW \citep{sohn13}. On such a high energy orbit, Leo I has proven to be an outlier as a tracer of the MW's mass thus far as it may or may not be gravitationally bound to the MW at present. In a follow up paper (Patel et al. 2017c, in prep.), we apply the Bayesian inference scheme to Leo I and show that there is a disparity in the inferred mass estimates of the MW that is similar to that of the LMC's such that the instantaneous method results in a much higher MW mass estimate compared to the momentum method. Therefore, implementing two different likelihood functions is not only a test of how reliable satellite properties are to make such inference measurements for the MW's mass, but it also separates the satellites that are on high energy first infall orbits from those which are much more circular and less energetic.

Ultimately, we conclude that the Bayesian technique utilising the observed angular momentum of satellites could be a powerful method for determining host mass by using a {\em population} of satellites belonging to the same halo, rather than focusing on individual cases. The momentum method differentiates between low and high orbital angular momenta and therefore could provide insight into the host mass based on the fraction of low and high angular momenta satellites it hosts. We will apply the momentum method to several low mass MW satellites in future work to determine whether more accurate and precise constraints for the MW's mass can be determined from the phase space information of nine MW satellites.

\section{Discussion}
\label{sec:discussion}

By computing the posterior probability distribution of the MW and M31's mass in several ways, we have explored how different orbital properties of massive satellites (LMC, M33) can provide insight on the most statistically significant halo mass of their hosts (MW, M31). Our first method takes the maximum circular velocity of the satellite, relative position, and relative velocity to determine the most probable halo mass of the MW and M31, respectively. By doing so, we find that the resulting halo mass distributions are fairly broad and in agreement with the current literature for both host-satellite systems, though they are biased by satellite orbital phase. By using satellite angular momentum as an input to the statistical scheme, we tend to find lower posterior mean values for the mass of the MW and M31 with slightly broader credible intervals (in log space). However, this method is more consistent as a function of time. 

In what follows, we discuss several caveats that are necessary to consider when combining high precision proper motions and cosmological simulations in a Bayesian statistical scheme. In particular, we focus on the bias introduced by the different orbital histories of the LMC and M33 individually. We also examine the specific case of M31's mass when we impose a close passage of M33 about M31 during the last few Gyr. Measurement error, cosmic variance error, and how they may affect the mass estimates of their hosts are also discussed. Finally, we compare our results to previous analyses.

\subsection{Orbital Histories of the LMC and M33}
\label{subsec:orbits}
\begin{figure*}
\begin{center}
\includegraphics[scale=0.6]{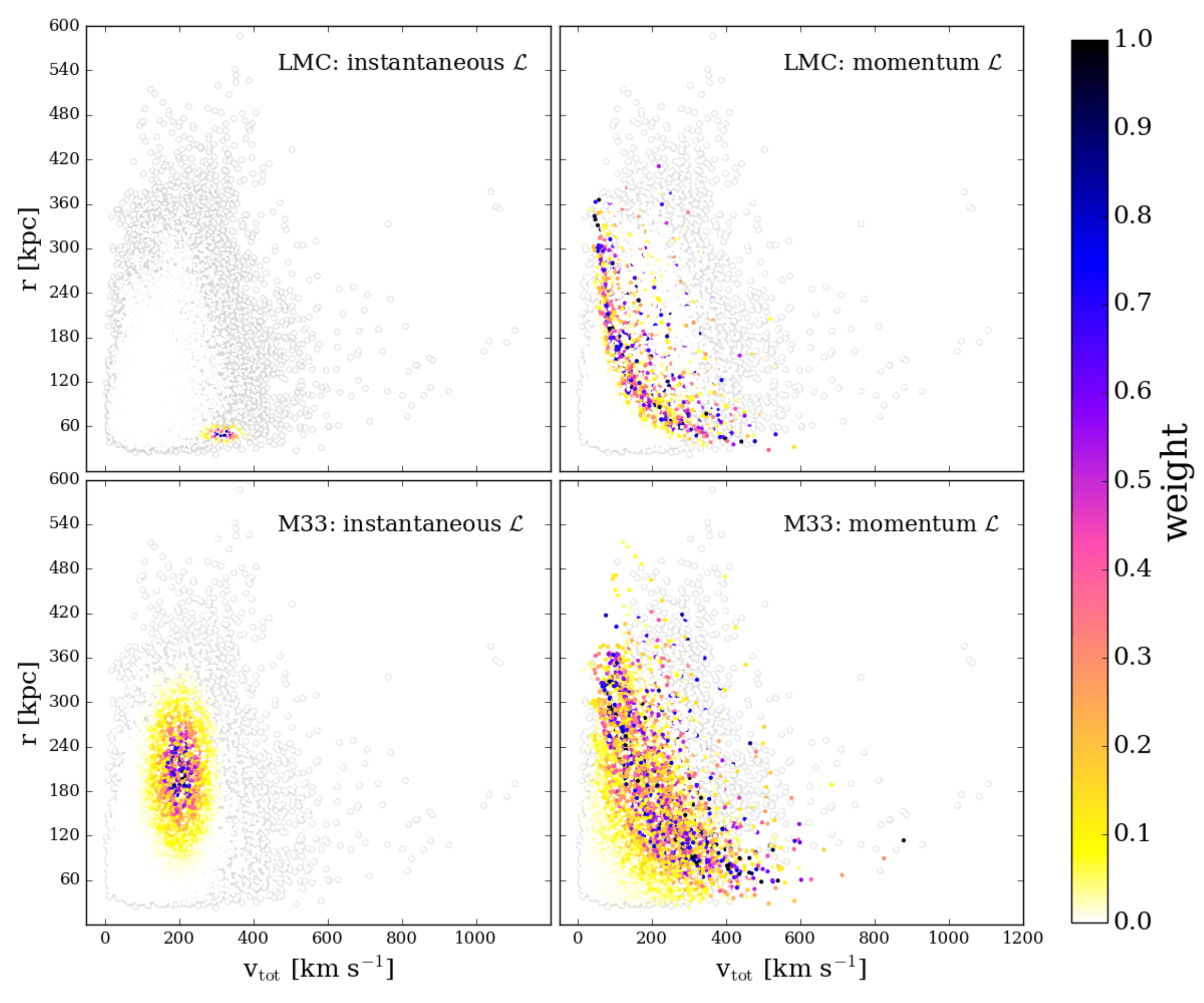}
\caption{ {\em Top:} The distribution of relative position and velocity for all Illustris-Dark host-satellite systems in the prior. The points are colored by the weight ($w_i$) assigned to each host halo, which represents how closely its associated subhalo resembles the current properties of the LMC. The weights are proportional to the respective likelihood functions and normalised here for easy comparison. The left panel shows the points colored by weights that are proportional to the instantaneous likelihood given by Eq.~\ref{eq:L2}. The right panel shows the same points now colored by the weights proportional to the momentum likelihood given by Eq.~\ref{eq:L3}. {\em Bottom:} The left and right panel are computed in the same fashion as the top row but now the points are weighted according to their similarity with M33's properties today. All points with a normalised weight < 0.025 have been colored white and are outlined in gray to easily distinguish between those host-satellite systems that are statistically significant to the inference scheme. The small fractions of colored points in the left column compared to the right demonstrates how significantly orbital phase can bias inferred host halo masses using the instantaneous method, as unique combinations of satellite position and velocity are less common in simulated analogs.}
\label{fig:posvelweights}
\end{center}
\end{figure*}

The LMC is on a significantly different orbital trajectory than M33--it is just past pericentre in its orbit, whereas M33 appears to be just past apocenter of a long period orbit or on first infall (see Paper I). Given its current orbital configuration, the LMC is about four times closer to the MW than M33 is to M31 and the LMC has a total relative velocity that is more than two times the velocity of M33 with respect to M31. Thus, the LMC is on a high energy orbit at present where its position and velocity exhibit extreme values. 

M33, however, is at a much more common place in its orbital trajectory in the context of massive satellite analogs. We found in Paper I that it appears to be approaching pericentre in the next few Gyr, so its current position and velocity are not rare compared to the positions and velocities of the satellites in the control sample, for example (see Fig.~\ref{fig:dynamicratios}). As a result, the posterior distribution of M31's halo mass is more broad compared to that of the MW. 

By using the orbital angular momentum of Illustris-Dark haloes instead of their positions and velocities, we find similarly broad posterior halo mass distributions for both the MW and M31. Orbital angular momentum is not exclusive to a unique combination of one position and velocity vector, but rather a set of positions and a set of velocities. Thus, it is not surprising that the resulting posterior distributions are more broad (and account for a larger fraction of phase space) than a posterior computed based on satellite position and velocity individually. In the case of the LMC, the two likelihood methods result in a noticeable difference in the shapes of the posterior PDFs, but for M33, both results are in good agreement with each other. We expect this is closely related to the rarirty of the LMC's current orbital configuration.

To demonstrate how significantly the orbital phase of a satellite galaxy affects the inferred mass of its host, Fig.~\ref{fig:posvelweights} shows the distribution of satellite position to satellite velocity for all host-satellite systems contained in the prior as a function of their likelihood weights for both the instantaneous and momentum likelihood functions. The weights (or color of the data point) in the top left panel are computed using the LMC's observed properties today (see Table~\ref{table:bayesparams}) and the instantaneous likelihood function, while the bottom left panel shows the same distribution calculated with the properties of M33 today. In all panels, the weights are normalised so that the results of both (left and right panels) methods can be easily compared.

Notice that the uniqueness of the LMC's position and velocity at present yields far fewer points with a non-zero weight in the instantaneous likelihood construction. As M33's position and velocity are somewhat more common amongst the phase space of massive satellite analogs, a larger fraction of points have non-zero weights from the instantaneous likelihood. The number of colored points in each panel of Fig.~\ref{fig:posvelweights} approximately corresponds to the ESS values listed in Table~\ref{table:ess}, which are representative of how many simulated haloes actually host subhaloes within the $\sim$2$\sigma$ average of the LMC/M33's observed properties.

The right column of Fig.~\ref{fig:posvelweights} is computed in a similar fashion, but uses the momentum likelihood construction given by Eq.~\ref{eq:L3}. Far more satellites in the prior exhibit orbital angular momenta similar to the LMC or M33, however, a more significant fraction of the prior aligns with M33's current orbital angular momentum instead of the LMC's. Again, this is a result of the LMC and M33 residing at different orbital phases in their trajectories about their respective host galaxies. Therefore, it appears that the disparity between the results of the two likelihood methods is much more drastic for satellites with a unique orbital configurations like the LMC.

\subsubsection{Bayesian Inference with a Close Pericentric Passage}
\label{subsubsec:closepassage}
\begin{table*}
\caption{The orbital sample descriptions for the criteria used in Paper I to quantify the plausibility of a recent, close passage of M33 about M31. The ARP sample is the subset of all orbits where M33 has made a pericentric passage about M31 during the last 6 Gyr. TI6 refers to the further subset of orbits where M33 also fell into the modeled halo of M31 during the last 6 Gyr. Finally, the strictest sample (RP100T) is designed to match the M33 orbital models suggested by \protect\citet{putman09} \protect\citep[see also][]{mcconnachie09}. The final column shows the average magnitude of specific orbital angular momentum for each sample. Here, the magnitude of the orbital angular momentum is computed using the initial M31-M33 position and velocity vectors for the orbital integrations. This table is adapted from Table 7 in Paper I. }
\label{table:m33table}
\centering
\begin{tabular}{c|c|c|c|c|c} \hline \hline
Identifier & N$\rm_{peri}$ & $\rm t_{peri}$ & $\rm t_{inf}$ & $\rm r_{peri}$ & avg. j$^{\rm obs}$  \\ 
& & [Gyr ago] & [Gyr ago] & [kpc]  & [kpc km s$^{-1}$]\\ \hline 
ALL & -- & -- & -- & -- & 27,656 $\pm$ 8,219\\ 
ARP & $\geq$1 &  $\leq 6$   & -- & --  & 23,094 $\pm$ 4,747 \\ 
TI6 & $\geq$1 &  $\leq 6$   &  $\leq$ 6  & --   & 22,113 $\pm$ 4,272\\ 
RP100T & $\geq$1 &  $\leq 3$   &  $\leq$ 6  & $\rm r_{peri} <$ 100 & 16,134 $\pm$ 1,118 \\ \hline
\end{tabular}
\end{table*}

\begin{figure*}
\begin{center}
\includegraphics[scale=0.6]{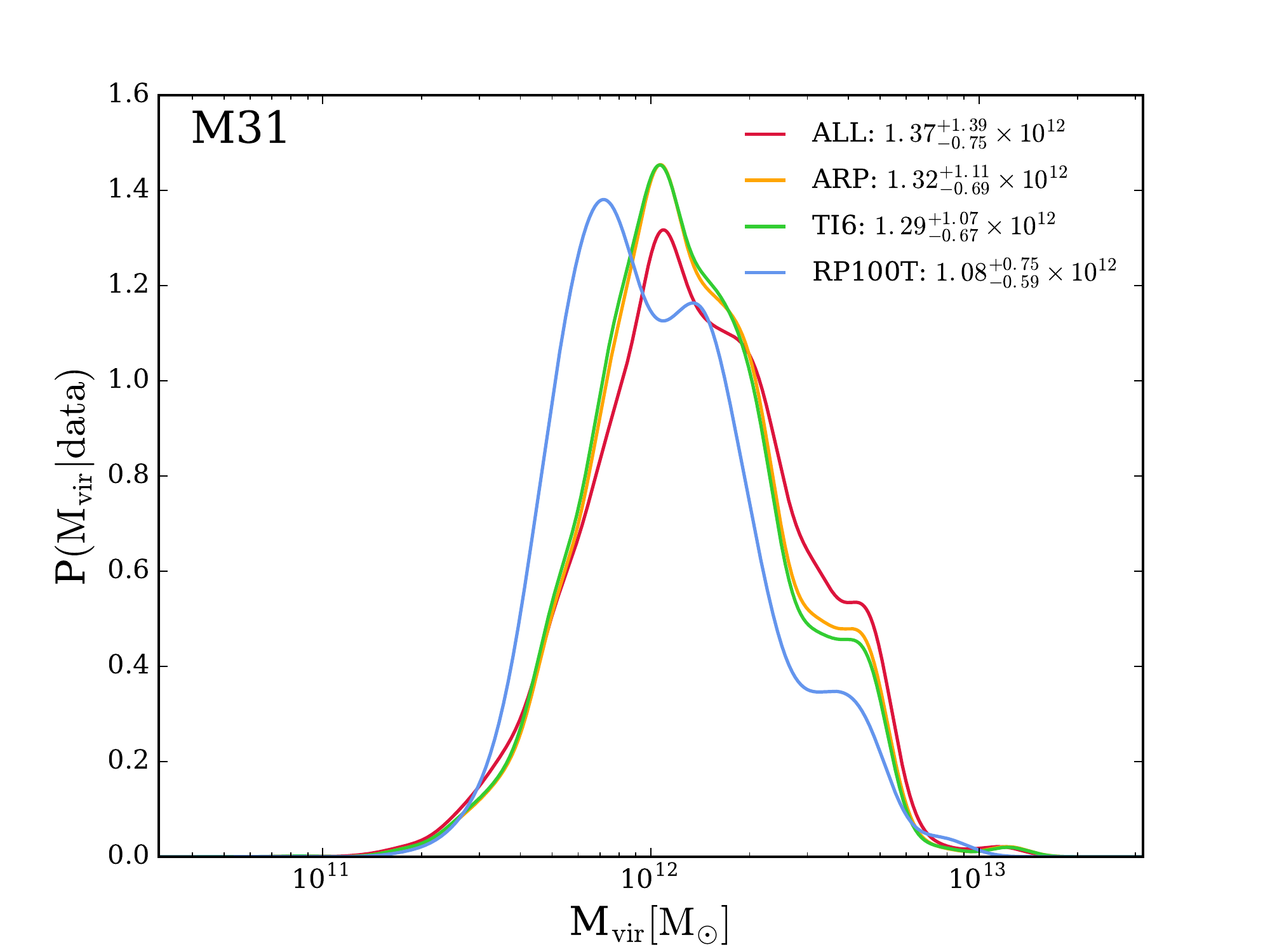}
\caption{The posterior distribution for M31's halo mass based on the properties of M33. Using the orbital angular momentum of M33 about M31 and its approximate $\rm v_{max}^{obs}$ value from Table~\ref{table:bayesparams}, we calculate the posterior distribution for M31's halo mass for each of the orbital samples described in Table~\ref{table:m33table}. The likelihood function (Eq.~\ref{eq:L3}) used to calculate importance weights from the prior takes the average orbital angular momentum and standard error for each sample as inputs. The RP100T criteria are designed to match the conventional M33 orbit involving a close (< 100 kpc), recent (< 3 Gyr) encounter between M31 and M33 \citep{putman09, mcconnachie09}, however, it yields the most discrepant M31 mass ($\sim10^{12}$ \Msun). This value contradicts estimates from the timing argument and abundance matching, further supporting the new M33 orbital models presented in Paper I.
\label{fig:m33orbit}}
\end{center}
\end{figure*}

In Paper I, we discussed the plausibility that M33 made a pericentric approach about M31 of < 100 kpc during the last 3 Gyr. This type of orbit is typically used to explain the formation of the stellar and gas disc warps observed in the structure of M33 today \citep{mcconnachie09, putman09}. We calculated 10,000 orbits for M33 in the allowed proper motion and velocity error space of the M31-M33 system to find that fewer than 1 per cent (RP100T sample) of all allowed orbits satisfy these criteria. When we further restrict our analysis to those orbits that achieved a pericentre < 55 kpc, we found that zero orbits satisfied this criteria. Table 7 of Paper I describes the exact orbital criteria applied to narrow down these statistics.

Here we revisit our M31-M33 analysis from Paper I to identify what M31 halo mass is preferred using the orbital angular momentum associated with the conventional and newly ascribed M33 orbital histories. By using the mean and standard error on the angular momentum of each individual orbital sample described in Table~\ref{table:m33table} (adapted from Paper I), we can estimate the mass of M31, applying the Bayesian methodology described in Section~\ref{subsec:angmom}. Therefore, momentum likelihoods are calculated for the prior such that the values of $\rm j^{obs}$ and $ \rm \sigma_j$ are adjusted to reflect each orbital criteria sample.

In Fig.~\ref{fig:m33orbit}, the posterior distribution in red, which refers to the full orbital sample (ALL), is the least restrictive and is identical to the black posterior distribution in the bottom panel of Fig.~\ref{fig:mombayes}. It encompasses all possible orbits of M33 in the allowed proper motion error space of the M31-M33 system. Each of the subsequent samples (orange, green, blue) requires one or more additional orbital parameters that further limits the value of the average angular momentum that goes into the momentum likelihood function. The ARP sample represents all orbits which show evidence of a pericentric passage in the last 6 Gyr. The TI6 sample is the subset of the ARP sample where M33 also fell into the halo of M31 in the last 6 Gyr. The most restrictive orbital criteria (RP100T) is constructed to match the orbits described in \citet{putman09} and \citet{mcconnachie09}, which both suggest that M33 made a close passage (< 100 kpc) about M31 in the last 3 Gyr.

In general, the posterior mean value for M31's mass shifts to lower values and the width of the credible intervals narrows as more criteria are added. Notice that the strictest criteria, the RP100T sample (blue), favors a posterior mean M31 halo mass $\sim10^{12}$ \Msun. This implies that if M33 really did achieve a distance of 100 kpc or less from M31 in the last $\sim$3 Gyr, then its corresponding orbital angular momentum suggests M31's mass would have to be quite low. We expect that the orbital angular momentum for an orbital sample that achieves a pericentre of 55 kpc or less would result in M31's mass being <$10^{12}$ \Msun. However, we cannot directly test this hypothesis since none of our 10,000 numerical orbits achieved these criteria in Paper I, so the corresponding orbital angular momenta for such an orbit is unknown. 

Our results are also supported by the cosmological analogs defined in Paper I. By extracting simulated orbital histories of massive satellite analogs that are capable of reaching such close pericentric distances recently, we found that they are likely to have host haloes with virial masses $\leq 1.5\times10^{12}$ \Msun. Generally, these cosmological and statistical results are also applicable to the MW-LMC system as the LMC is currently just past pericentre ($\sim$ 50 kpc) in its orbital history, thereby suggesting that the MW's mass should also be $\sim$1-1.5$\times10^{12}$ \Msun.

We conclude that the inferred M31 virial mass of $\sim$$10^{12}$ \Msun is well below the current expectation for the M31 mass based on abundance matching \citep{moster13}, the timing argument (vdM12), and satellite kinematics \citep{watkins10}. These low M31 mass results reinforce the assertion in Paper I that M33 is unlikely to be on an orbit that yields a recent, close encounter with M31, contrary to conventional wisdom. These results also further illustrate that the statistical methods applied here could be used to assess the plausibility of a given orbital history for a satellite. As we have shown for the M31-M33 system, the resulting halo mass of M31 is at odds with observational evidence, suggesting a a revision of M33's conventional orbital model.

 \subsection{Measurement Errors on the Observed Properties of the LMC and M33}
 \label{subsec:observederrors}
A major source of error to consider for Bayesian inference schemes such as the one outlined in this paper is the error due to the measurement error, or the precision with which observational data were measured. The measurement errors on the observed properties of satellites can significantly impact the resulting host halo masses such that the posterior means shift and/or the credible intervals narrow and shift as measurement precision increases. Some aspects of this were already demonstrated in Section~\ref{sec:results} when several velocity and orbital angular momenta values of M33 were used to estimate the mass of M31.

Here, we recompute the results from Sections~\ref{subsec:kinematics} and \ref{subsec:angmom} with measurement errors that are only half as large as those listed in Table~\ref{table:bayesparams} to quantify how much precision affects our results. By doing so, we aim to illustrate that this technique may be a powerful method moving forwards as future measurements are made with smaller uncertainties. The measurement errors on M33's kinematics and dynamics are naturally larger given the distance of M33 from the MW compared to the LMC. 

Using the instantaneous likelihood, we find that the resulting posterior mean masses for the MW and M31 are: $1.70^{+1.59}_{-0.49}\times 10^{12}$ \Msun, $1.68^{+1.03}_{-0.75}\times 10^{12}$ \Msun (68 per cent credible interval). The mean mass of the MW changes insignificantly since the LMC's position and velocity are known to within 10 per cent, but the 68 per cent credible interval increases slightly. M31's posterior mean halo mass increases by 16 per cent when the measurement errors on M33's position and velocity are halved. The 68 per cent credible interval's upper limit shifts up as the measurement error on total velocity decreases from 38 \kms to 19 \kms. Reducing the measurement error on M33's position by half still amounts to about 24 kpc, therefore, the posterior distribution is still quite broad. 

Using the momentum likelihood method, we find that the posterior mean virial masses within the 68 per cent credible intervals are $1.06^{+0.76}_{-0.60}\times 10^{12}$ \Msun (MW) and $1.60^{+1.49}_{-0.82}\times 10^{12}$ \Msun (M31). Again, the MW's mass estimate changes minimally and M31's inferred mass goes up by about 16 per cent as the measurement errors are halved. Therefore, measurement error does contribute to the overall uncertainty of this method. However, its contributions are minor and even $\leq$ 10 per cent precision on M31 and M33's distance and proper motions may not provide an extremely precise range for M31's halo mass.

\subsection{Cosmic Variance and its Effect on the Mass Estimates of the MW and M31}
The posterior means and credible intervals, which summarize the posterior PDFs calculated in this work, simultaneously factor in both the measurement error on the observed satellite properties and the irreducible uncertainty associated to the imperfect correlation between host halo masses and satellite dynamics. The latter is often referred to as `cosmic variance'. In other words, even if the 6D phase space information for all satellites of the MW (or M31) was known and if that information was incorporated into our importance sampling technique, there would still be an intrinsic scatter due to the cosmology of the Illustris-Dark simulation, as it is only one realization of the universe.

The magnitude of measurement error and how it affects the posterior PDF was already discussed in Section~\ref{subsec:observederrors}. Here, we wish to quantify the remaining uncertainty, if any, due to cosmic variance. This irreducible uncertainty is essentially the variance on the conditional probability distribution $\rm P(M_{vir}|\, {\bf x}$). Even if we knew the observational data $\rm {\bf d}$ perfectly with zero measurement error, there would still be some intrinsic scatter associated with $\rm P( M_{vir}|\, {\bf x}$) due to the correlation between host halo mass and satellite dynamics. (i.e. $\rm {\bf x}$ does not perfectly predict \Mvir). We do not have an analytic function to describe $\rm P( M_{vir}|\, {\bf x}$), but we do have samples from $\rm P(M_{vir}, {\bf x}$), which we can use to quantify the magnitude of cosmic variance error. 

Our method for computing posterior PDFs requires us to assume a finite measurement error so that our ESS is reasonably large. In practice, we can treat a random set of host-satellite systems from the prior as the {\em data} (assuming some fixed measurement error) and apply our two likelihood methods. By doing so, we measure how well our statistical method can predict the true host halo mass and if the contribution from this irreducible uncertainty is properly accounted for by the reported credible intervals. 

For this purpose, we randomly select 25 host-satellite systems from the prior where the satellite's position relative to its host is < 150 kpc. We have chosen this distance so that we can apply reasonable measurement errors that are informed by the true properties and measurement errors for satellites in the MW's halo. Therefore, the measurement errors assigned to $\rm v_{max}, r, v, and \, j$ are 10, 10, 15, and 18 per cent, respectively.  

For all 25 systems, we calculate the host halo mass using both the instantaneous and momentum methods, ensuring that the ESS remains reasonably large compared to our bootstrapping results in Section~\ref{sec:results}. We then compute the root mean square (RMS) error of the posterior log halo mass estimate relative to the true log halo mass across our 25 host-satellite test cases, once for each likelihood method. The posterior log mass estimate is determined from the total posterior pdf using either the instantaneous or momentum method.% The RMS error here is 

For the instantaneous method, the ratio of the RMS error across all 25 systems to the average of the posterior standard deviations of $\log_{10}$\Mvir ($\rm \sigma_{post}$) is approximately 0.78\footnote{Note that posterior standard deviation refers the standard deviation of the total posterior PDF computed for each given system (i.e.for the MW-LMC, this would be the standard deviation of the black curves shown in the top panels of Figs.~\ref{fig:instbayes} or \ref{fig:mombayes}.)}. For the momentum method, we find this ratio is about 0.87. Therefore, both the instantaneous and momentum methods accurately encompass and even overestimate the uncertainty due to cosmic variance since the average of the standard deviations is always greater than the RMS error. The RMS errors, averages of the posterior standard deviations (in $\log_{10}$\Mvir), and the ratio of these quantities are listed in Table~\ref{table:errors}. These quantities are reported on a log scale to avoid any bias from transforming posterior quantities calculated in log space to linear space. 

If we reduce the assigned measurement errors listed above by half (i.e. 5, 5, 7.5, 9 per cent) and redo our analysis for all 25 systems, we find that the RMS errors and the average posterior standard deviations for each respective method change insignificantly or remain the same. Little to no change in these quantities demonstrates that the measurement error is not the main source of uncertainty. Instead, the intrinsic scatter related to cosmic variance and therefore the imperfect correlation between host halo mass and satellite dynamics is the key source of uncertainty. Thus, more accurate halo mass estimates are not necessarily guaranteed with these methods if higher precision proper motion and distance measurements are obtained for a single satellite, but including measurements of more than one satellite galaxy may result in better halo mass constraints. This will be the focus of future work.

\begin{table}
\centering
\caption{The first two columns give the RMS error and the average posterior standard deviations ($\rm \sigma_{post}$) in $\rm\log_{10}$\Mvir across 25 host-satellite test cases randomly chosen from the prior. The final column shows the ratio of these quantities. The first two rows indicate the values using the instantaneous likelihood function where the assigned measurement errors (ME) on $\rm v_{max}, r, v, and \, j$ are respectively 10, 10, 15, and 18 per cent and then reduced by half to 5, 5, 7.5, and 9 per cent. The last two rows show the same quantities for the momentum likelihood functions across the same 25 test cases.}
\label{table:errors}
\begin{tabular}{c|c|c|c}
& RMS & avg. $\rm \sigma_{post}$ & $\rm \frac{RMS}{avg.\,\sigma_{post}}$ \\ 
& [dex] & [dex] & \\ \hline
Instantaneous & 0.20 & 0.26 & 0.78 \\ 
(10, 10, 15,18 per cent ME)  & & & \\ \hline
Instantaneous  & 0.20  & 0.24 & 0.84 \\ 
(5, 5, 7.5, 9 per cent ME) & & & \\ \hline \hline
Momentum & 0.27 &  0.30 &  0.87\\ 
(10, 10, 15,18 per cent ME)  & & & \\ \hline
Momentum  & 0.27 & 0.30 & 0.91\\ 
(5, 5, 7.5, 9 per cent ME) & & & \\ \hline \hline
\end{tabular}
\end{table}

\subsection{Comparison to Previous Work}
\label{subsec:previouswork}
\subsubsection{The Mass of the MW}
Using the kinematics of the LMC and SMC derived from \citep{k06a,k06b}, B11 finds the virial mass of the MW is $1.2^{+0.7}_{-0.4}\times10^{12}$ \Msun within the 68 per cent credible interval. G13 estimated the mass of the MW based on the properties of the MCs (from K13) and the larger Local Group environment to find a MW mass of $\rm \log M_{200}=12.06^{+0.31}_{-0.19}$ encompasses the 90 per cent credible interval.

These results were computed by applying a statistical inference scheme to the combination of the Bolshoi cosmological simulation and the observed properties of the MCs, with respect to the year in which these studies were conducted. The Bolshoi simulation has a much larger volume (nearly 37 times larger) than the simulation used in this analysis, Illustris-Dark. Secondly, the simulations use slightly different cosmological parameters. Finally, we choose our priors using different selection criteria. Together these differences account for the variation between our MW mass results. 

Due to the significantly smaller volume of Illustris-Dark, we find that it is actually statistically impossible to apply the exact G13 (or B11) methodology to infer the MW's virial mass. When we choose our priors identically to theirs, which requires that each host halo has an analog of the LMC and the SMC, we find that zero systems lie within the average $2\sigma$ range of the observed properties of the MCs from a redshift of $z=0.26$ to $z=0$. With no matches to the observed properties of the MCs, the ESS is effectively zero and therefore the importance sampling technique cannot be applied with two massive satellites akin to the MCs in Illustris-Dark. However, as we demonstrated in Section~\ref{sec:results}, requiring only one massive satellite analog when selecting the prior provides a reasonable statistical sample in Illustris-Dark for which importance sampling can be accomplished. 

Therefore, by modifying the analysis of B11 to use only the kinematics of one massive satellite (the K13 properties for the LMC), we find that the virial mass of the MW is $\rm \log_{10}M_{vir}=12.23^{+0.25}_{-0.16}$ \Msun within the 68 per cent credible interval and $\rm \log_{10}M_{vir}=12.23^{+0.43}_{-0.43}$ within the 90 per cent credible interval. Our posterior mean is consistent with G13's result when their value of $\rm M_{200}$ is extrapolated to approximately \Mvir. The small discrepancy between posterior means and the width of the credible intervals can likely be attributed to using just one satellite in our analysis. The G13 posterior mean is slightly lower than our findings, and we suspect this might be driven by the inclusion of the SMC and its low velocity relative to the MW. The difference between our results and B11's is mainly driven by the prior selection criteria, and the subsequent inclusion of both MCs. B11 adopts the properties of the MCs that were derived from the \citet{k06a, k06b} proper motions, which not only changed significantly in K13, but also have much higher measurement errors than the revised values of K13.

Finally, we have compared the halo mass functions for the Bolshoi simulation and Illustris-Dark, and we conclude that the choice of cosmology does not contribute significantly to the difference in inferred MW masses from our analysis and G13's (or B11's). If we adopt a fiducial mass for the LMC that is 100 times less than the MW's, then by comparison of the abundance ratio of haloes with a mass of $\sim$1.7$\times10^{12}$ \Msun (our instantaneous MW result) and $\sim$1.7$\times10^{10}$ \Msun we can assess the magnitude of error introduced by different values for $\sigma_8$ and $h$. In the Bolshoi simulation, we find an abundance ratio of 66.1 and in Illustris-Dark, the ratio is 66.6. Thus, the simulations agree to within 1 per cent of each other for halo abundances and using one over the other would not affect our MW mass estimate.

\subsubsection{The Mass of M31}

Recent mass estimates for M31 are directly affected by the assumed value of its transverse motion and corresponding measurement error. Until the proper motion of M31 was directly measured by S12, the assumed values spanned a generous range of velocities. While no previous authors have applied a Bayesian scheme using satellite dynamics to infer the mass of M31, as we have done in this work, our results are still in good agreement with estimates resulting from several independent techniques. We highlight several selected results below.

\cite{fardal13} used N-body models that reproduce the Giant Southern Stream in the halo of M31 to estimate its enclosed mass as $\rm log_{10}M_{200}=12.32\pm0.1$. \citet{watkins10} used the line of sight velocities and distances to 23 satellite galaxies in the halo of M31 to find a mass of $\rm M_{300} = 1.4\pm0.4\times10^{12}$ \Msun. This is approximately equivalent to the virial mass and is in very good agreement with our mass estimates even though these results were derived using multiple satellites and we only use the properties of one satellite galaxy in the halo of M31 (i.e. M33). 

Many authors have used the well known timing argument to estimate the masses of the MW, M31, and the Local Group simultaneously. Some recent works include that of vdM12, who estimate M31's virial mass to be $\sim$1.5-1.7$\times10^{12}$ \Msun for a low tangential velocity and nearly radial orbit relative to the MW. More recently, \citet{carlesi16} estimated the mass of M31 using the timing argument in a Bayesian fashion using $\Lambda$CDM cosmological simulations and find values of 1-2$\times10^{12}$ \Msun. These studies incorporate the measured tangential velocity of M31 in their models and are therefore most similar to our analysis. Note, however, that these mass estimates contradict the rather low M31 mass ($\sim$$10^{12}$ \Msun) inferred by imposing a close M31-M33 encounter (Section~\ref{subsubsec:closepassage}).

While our mass estimates for M31 using the observed properties of M33 are no more profound than previous estimates, they do help test the viability of different likelihood functions. As the full 6D phase space information from future proper motion measurements of other M31 satellites become available, this hypothesis can be tested.

\section{Conclusions}
\label{sec:conclusions}
We have modified and expanded the Bayesian inference scheme developed by B11 to infer the masses of the MW and M31 using the observed properties of their satellites. This method combines high precision astrometric measurements of satellites with high mass resolution cosmological simulations in a statistical fashion to constrain host galaxy mass. We find this to be a promising statistical scheme to learn about the hosts of satellites in the era of high precision astrometry. 

By analyzing a set of massive satellite galaxy analogs (i.e. analogs of the LMC and M33) defined in Paper I, we confirmed that orbital angular momentum is well conserved over time and is therefore an ideal orbital property for constraining the larger host environment of a satellite galaxy. We therefore expand the B11 inference scheme by creating a new likelihood function that uses orbital angular momentum instead of the individual position and velocity of a satellite relative to its host to infer the mass of the MW and M31, respectively. Therefore, the masses of the MW and M31 are each determined using haloes from the Illustris-Dark cosmological simulation as the prior in the following ways: (1) apply a likelihood function that uses the current position and velocity (instantaneous method) of the LMC or M33 to determine host halo mass and (2) apply a likelihood function that uses current orbital angular momentum (momentum method) of the LMC or M33 to infer host halo mass. 

Since the instantaneous method uses satellite position and velocity, which are both susceptible to large variations at different orbital phases (i.e. pericentre vs. in between apo- and pericentre), overall, it is a less reliable method for determining host halo mass with cosmological analogs from Illustris-Dark. Instead, we find that the momentum method produces more accurate though less precise results compared to the instantaneous method. 

The results of the instantaneous method are $\rm \log_{10}M_{vir}=\left[12.23^{+0.25}_{-0.16}\, (MW), 12.16^{+0.27}_{-0.28}\,(M31)\right]$, suggesting that the MW is more massive than M31, contradicting conventional wisdom. The new likelihood function developed in this paper, the momentum likelihood, yields the following mass estimates: $\rm \log_{10}M_{vir}=\left[12.01^{+0.25}_{-0.34} (MW),\, 12.12^{+0.32}_{-0.35}\,(M31)\right]$, where M31 is now more massive than the MW. 

Furthermore, when we require M33 analogs to have a made a recent (< 3 Gyr ago), close encounter (< 100 kpc) relative to its host halo, our statistical analysis yields an estimated M31 mass of only $\sim$$10^{12}$ \Msun. Such a low mass is inconsistent with several independent M31 mass estimates, and therefore further supports the new orbital histories for the M31-M33 system presented in Paper I. These results also imply that such statistical methods may be useful in constraining satellite orbital histories such that the imposition of incorrect orbital trajectories might result in unlikely host halo masses, and therefore could help constrain the plausibility of a given orbital scenario (e.g. the case of M33). Our cosmological analogs in Paper I preferred an M31 mass $\geq1.5\times10^{12}$ \Msun based on the orbital energy of M33 and a majority of these analogs are also on a first infall orbit. Furthermore, our numerical orbit integrations independently showed that a first infall scenario was very plausible in the proper motion error space of M31-M33, demonstrating the links between satellite dynamics, host mass, and orbital histories.

We have also shown that the instantaneous method is more susceptible to bias as a function of time and orbital history by applying it along the LMC's past orbital trajectory, which was calculated in Paper I. When the LMC's time-dependent position and velocity are used as inputs for the instantaneous method, the inferred MW masses over time show deviations of approximately a factor of two. In contrast, the momentum method infers consistent MW masses over time, regardless of the LMC's orbital phase. Therefore, the combined analysis of the MW-LMC and M31-M33 system at present and the application of the statistical scheme as a function of time together demonstrate that the momentum method is the most reliable method for estimating host halo mass for a variety of host-satellite systems. 

A close inspection of sources of error that may contribute to our statistical method has demonstrated that the precision and accuracy of mass estimates for the MW and M31 are primarily dominated by the irreducible uncertainty caused by cosmic variance. While our methods correctly capture the magnitude of this uncertainty, higher precision measurements of proper motions and distances to a single massive satellite galaxy may not guarantee better measurements of its host's mass. However, {\em simultaneously} incorporating precise measurements of more than one satellite (i.e. a population of satellites) in a given host halo may improve our overall mass estimates.

While this work has only used the dynamical information of one satellite galaxy in each of the MW and M31's haloes, proper motions are currently available for nine other dwarf satellite galaxies of the MW besides the MCs, and many more will become available in the near future. Now that we have established that estimating the mass of the MW should be approached from the perspective of orbital constants, we must continue to improve our statistical methods such that the maximal amount of phase space information for satellites (and eventually other substructures in the MW and M31's haloes) can be used to achieve high precision mass measurements of the MW and M31. 

In our next paper (Patel et al. 2017c, in prep), we will calculate the MW's mass using the properties of each low mass dwarf satellite (derived from their 6D phase space information), and finally, we will compute the MW's mass using the combined information from the ensemble of dwarf satellites. By doing so, we aim to illustrate the full power of this technique in the era of high precision astrometry. 

Interestingly, this technique can be modified to address several broader topics in near-field cosmology. We have already established that it is trivial to add more satellites into consideration for the MW's halo and plan to demonstrate this in upcoming work. When proper motion data becomes available for M31 satellites (e.g. HST-GO proposal \#14769 for NGC 185/147), this technique will be easily applicable to the M31 system. Furthermore, one could extend the statistical method that we have outlined here to include not only the magnitude but also the direction of satellite specific orbital angular momentum vectors (e.g. to address the alignment of satellite orbits). Finally, a prior sample chosen from a suite of cosmological zoom simulations could be used so that both globular clusters and dwarf satellite galaxies can be included in this type of analysis.

\section*{Acknowledgments}
EP is supported by the National Science Foundation through the Graduate Research Fellowship Programme funded by Grant Award No. DGE-1143953. KM is supported at Harvard by NSF grants AST-1211196 and AST-156854. This research was also funded through a grant for HST programme AR-12632. Support for AR-12632 was provided by NASA through a grant from the Space Telescope Science Institute, which is operated by the Association of Universities for Research in Astronomy, Inc., under NASA contract NAS 5-26555. Orbital model calculations were performed with the El Gato cluster at the University of Arizona, which is funded by the National Science Foundation through Grant No. 1228509. The Illustris simulations were run on the Odyssey cluster supported by the FAS Science Division Research Computing Group at Harvard University. We thank Vicente Rodriguez-Gomez for the \texttt{SUBLINK} merger trees and the Illustris collaboration for making their data and catalogs public. We are also grateful to Roeland van der Marel, Dennis Zaritsky, Tony Sohn, Risa Wechsler, Mark Fardal, Hans-Walter Rix, and Yao-Yuan Mao for useful discussions that have contributed to this paper.

{\em Software:} The code developed for this project is available from \url{https://github.com/ekta1224/BayesToolsSatDynamics} under the MIT open-source software license. This research also utilised: \texttt{IPython} \citep{ipython}, \texttt{numpy} \citep{numpy}, \texttt{scipy} \citep{scipy}, and \texttt{matplotlib} \citep{matplotlib}.



\bibliographystyle{mnras}
\bibliography{myrefs}

%%%%%%%%%%%%%%%%% APPENDICES %%%%%%%%%%%%%%%%%%%%%

\appendix

\section{The Stability of Maximum Circular Velocity}
\label{sec:appendixA}

\begin{figure}
\centering
\includegraphics[scale=0.55]{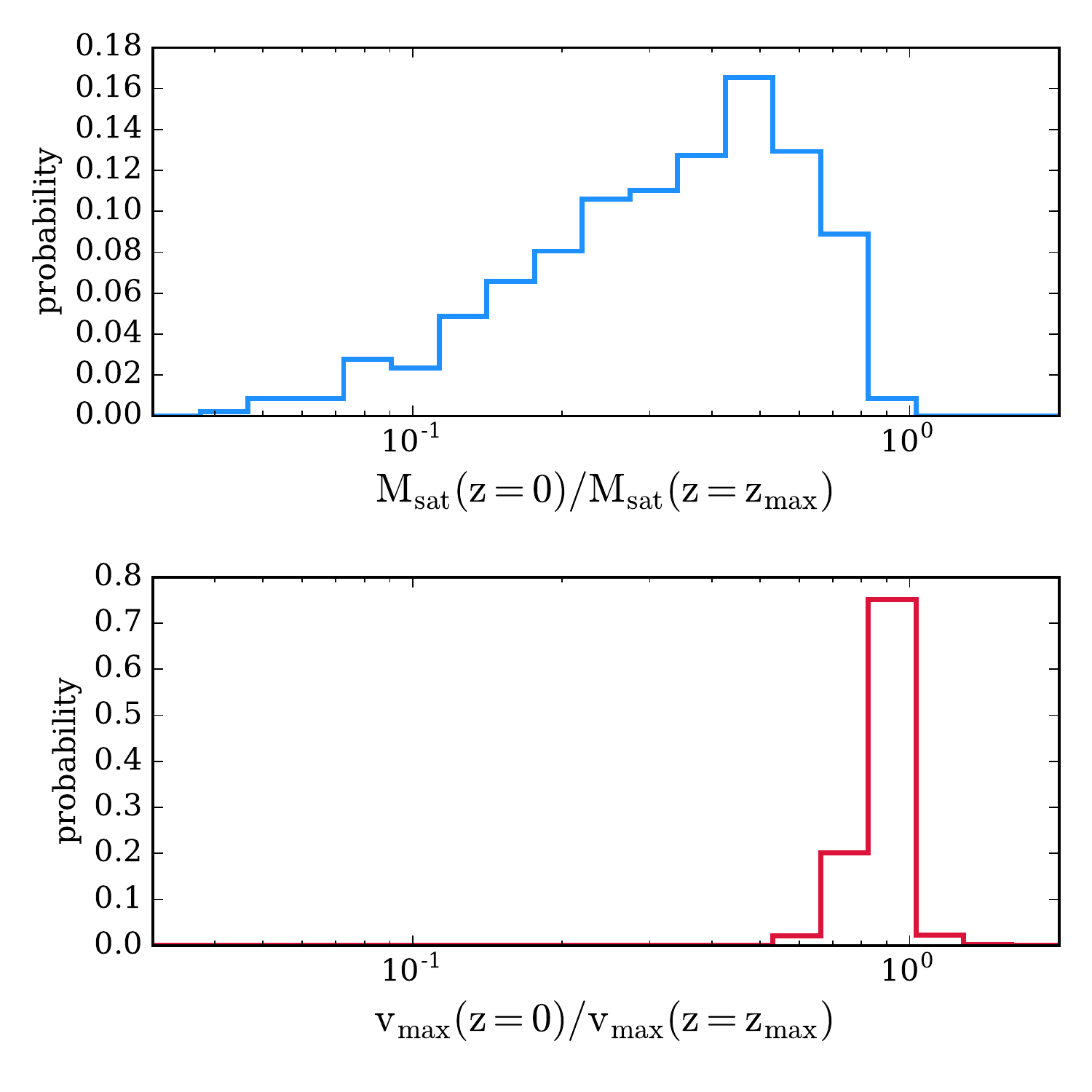}
\caption{{\em Top:} The distribution of the ratio of satellite mass at $z=0$ to satellite mass at maximal mass for the Illustris-Dark control sample of massive satellite analogs. {\em Bottom:} The distribution of the ratio of satellite maximal circular velocity at $z=0$ to satellite maximal circular velocity at the time of maximal mass, $\rm z_{max}$. The latter shows much less variation for the overall massive satellites analog sample when the quantities are binned over the same range and with a fixed bin width. The y-axis denotes the probability of a specific ratio.
\label{fig:vmax}}
\end{figure}

The maximum circular velocity of a dark matter halo is generally more stable than the subhalo mass (typically defined as the sum of bound particle masses) as a function of time. It is well known from numerical simulations that dark matter haloes lose mass from their periphery rather than their most central region due to tidal disruptions \citep{klypin99}. Secondly, much of the fluctuation in subhalo mass history is caused by the numerical effects of halo finding algorithms themselves. The friends of friends \citep[FoF,][]{davis85} algorithm and \texttt{SUBFIND}, the substructure identification code used in Illustris-Dark are both sources of mass bias for subhaloes. The identification of haloes and assignment of substructures within individual haloes therefore raises concerns for analyses using several, consecutive simulation snapshots. 

Our Bayesian inference methods consider all subhaloes in a window of $z=0-0.26$, or 20 snapshots of the Illustris-Dark output. For this study, we choose $\rm v_{max}$ as a proxy for subhalo mass and identify analogs of the LMC and M33 purely based on $\rm v_{max}$. Both galaxies have well-defined rotation curves \citep{vdmnk14, corbelli03}, motivating our choice to move away from subhalo mass and its related uncertainties. This choice is also justified from evidence in Paper I, which shows that at least 49.36 per cent of the massive satellite sample in Illustris-Dark make one pericentric passage within 100 kpc of its host's centre of mass recently (within the last 3 Gyr). At such separations, tidal stripping will remove material (the tidal radius of the LMC for example is expected to be between 20-30 kpc) and will likely cause significant mass and radius fluctuation for massive satellite analogs. 

Fig.~\ref{fig:vmax} compares the change in circular velocity and subhalo mass at two epochs in the lifetime of the Illustris-Dark massive satellite analogs control sample (see Section~\ref{subsec:control}). The ratio is computed using the $z=0$ properties compared to the epoch at which the satellites reach their maximal mass, $\rm z_{max}$. This epoch is identified using the \texttt{SUBLINK} merger trees \citep{rg15}. Fig.~\ref{fig:vmax} illustrates that $v_{\rm max}$ is generally constant over a few Gyr timescale while $\rm M_{\rm sat}$ is more variable. The distributions are plotted for a fixed bin width to emphasize the range of each distribution. In a fixed period of time, the subhalo mass can decrease by up to a factor of 10, whereas the circular velocity remains consistent within a factor of a few at most. The subhaloes with highest mass loss ratios generally have a time of maximal mass $\geq$ 3 Gyr ago, which is about half of an average orbital period. 

Note that in Paper I and prior to Section~\ref{sec:bayesian} in this work, massive satellite analogs are chosen primarily based on mass provided that they survive until $z=0$ within the virial radius of their hosts. Since Paper I tracks the dynamical histories of massive satellite analogs, the evolution of subhalo mass across time did not affect our conclusions. In this work, $\rm v_{max}$ allows us to choose a consistent sample of massive analogs across 20 snapshots in Illustris-Dark.
 
\section{Kernel Density Estimation for Bayesian Inference}
\label{sec:appendixB}

Histograms are a common way to represent posterior probability distributions. This process typically goes as follows: choose a mass bin for which to compute the posterior of the sample at that mass range and repeat for multiple, contiguous mass bins until you have computed enough data points to form an informative distribution. This method yields one point per mass bin where the point represents the total probability for the set of samples only in that host mass range. However, calculating the posterior for a finite number of bins will not finely sample the posterior well and can be computationally expensive. Consequently, the summary statistics (i.e. mean and credible intervals) over the set of samples become difficult to compare directly with a coarsely sampled host halo mass probability distribution. 

One way to sample the posterior more finely is to compute the probability of the target parameter (i.e. host halo mass) by taking bins in the mass range of interest through a gaussian kernel density estimation (KDE) technique. Kernel density estimation allows us to smooth over the posterior PDF to avoid harsh edges caused by a coarse sampling of the grid over the target parameter range, as found in the histogram method. 

In the KDE method, we create a uniformly spaced grid over a reasonable range for the host halo mass exhibited by the haloes in the prior. Each halo in the prior is represented on this mass grid by a Gaussian distribution centered at the halo's mass with a standard deviation given by the optimal bandwidth determined by the whole sample.  We scale each halo's Gaussian by its normalised importance weight, and then sum these distributions over all the halos.  This results in a smooth representation of the posterior PDF in mass, as we have shown in Figs.~\ref{fig:instbayes} and~\ref{fig:mombayes}. The rule of thumb for choosing the optimal bandwidth, $h$, is determined by \cite{silverman98} as 
\begin{equation}\rm h=\left(\frac{4\sigma^5}{3n}\right)^{1/5}\approx 1.06\sigma n^{-1/5}. \label{eq:bandwidth} \end{equation}
Here, $\sigma$ is the posterior standard deviation estimated from the importance-weighted halo masses and $n$ is the sample size where each of the values is typically given a weight ($w_i$) of one. However, in importance sampling, each sample is not given an equal weight, thus $n$ must be substituted with the effective sample size \citep[][ESS]{kong92}, 
\begin{equation}\rm  ESS= \frac{(\sum_{i=1}^m w_i)^2}{\sum_{i=1}^m w_i^2}. \label{eq:ess} \end{equation}
To preserve the machine precision of the importance sampling technique, all weights should be calculated and stored as $log(w_i)$ and exponentiated when used in the calculations of bandwidth, ESS, and summary statistics. It should be noted that regardless of technique, mean values and credible intervals should always be calculated over the full set of samples, not from the binned or KDE results. 

%%%%%%%%%%%%%%%%%%%%%%%%%%%%%%%%%%%%%%%%%%%%%%%%%%
% Don't change these lines
\bsp	% typesetting comment
\label{lastpage}
\end{document}